\documentclass[a4paper,11pt]{article}
\usepackage{jheppub} 
\usepackage{lineno}
\usepackage{braket}

\title{Expectation values after an integrable boundary quantum quench}

\author{Zoltán Bajnok, Dávid Fülepi, Máté Lencsés}
\affiliation{HUN-REN Wigner Research Centre for Physics, 1121 Budapest,
Konkoly-Thege Miklós út 29-33, Hungary}

\abstract{We investigate an integrable boundary quench, in which one integrable boundary condition is suddenly switched to another. We develop a general framework for analyzing the resulting real-time dynamics based on form factors of bulk and boundary-changing operators. We first study the problem at the conformal point of the Lee–Yang model and then extend the analysis to its massive perturbation, where we examine the time evolution of the pre-quench vacuum and compute the vacuum-to-vacuum matrix elements of local operators inserted after the quench. The analytical results are validated by numerical calculations using the truncated conformal space approach adapted to boundary-changing situations.}

\begin{document}
\maketitle
\flushbottom

\section{Introduction}\label{sec:intro}

Quantum quenches provide a paradigmatic framework for investigating non-equilibrium dynamics in extended quantum systems. They have been widely employed to study relaxation and thermalization phenomena ~\cite{Deutsch1991PhysRevA.43.2046,Srednicki19941994PhRvE..50..888S}, both experimentally~\cite{Kinoshita2006Natur.440..900K,2007Natur.449..324H,2012Sci...337.1318G,2015Sci...348..207L} and theoretically~\cite{2006PhRvL..96m6801C,2011PhRvL.106v7203C}. For a comprehensive overview, see Ref.~\cite{2018JPhB...51k2001M}.

In a typical quench protocol, a system is initialized in the ground state of a Hamiltonian (or, more generally, in a specified pre-quench state), after which a parameter of the Hamiltonian is abruptly changed at time $t=0$. When the perturbation is applied uniformly across the system, the process is referred to as a \textit{global quantum quench}. Such protocols are analytically tractable in certain cases, notably in (1+1)-dimensional conformal field theories~\cite{2006PhRvL..96m6801C} and when the pre- or the post quench Hamiltonian is integrable~\cite{2014JPhA...47N2001D,2017JPhA...50h4004D,2012JSMTE..04..017S,sineGquench2014JSMTE..10..035B,2018ScPP....5...27H}.

 In contrast to global quenches,  one may also consider \textit{local quenches}, where the perturbation is spatially confined, such as changing a local bond interaction~\cite{2007JSMTE..06....5E,2014JSMTE..03..023I}, or applying a single spin flip in a spin chain~\cite{Fronts2024ScPP...16..138K}. Other local non-equilibrium protocols include, for example, the joining of two semi-infinite subsystems, which, in the conformal case can be effectively described by conformal field theory techniques~\cite{2007JSMTE..10....4C,2011JSMTE..08..019S,BernardDoyon1,BernardDoyon2,BernardDoyon3,BernardDoyon4}, and within the framework of generalized hydrodynamics in integrable models~\cite{GHD12016PhRvX...6d1065C,GHD22016PhRvL.117t7201B}. 
Situations with inhomogeneous post-quench Hamiltonian and inhomogeneous initial states have been also considered in~\cite{2020NuPhB.95415002D} and~\cite{2022ScPP...12..144H}, respectively.

A unifying physical picture emerges across these scenarios: the quench generates quasi-particle excitations that propagate through the system with bounded velocities. This leads to the characteristic light-cone spreading of correlations and information~\cite{CardyCalabresePairs2005JSMTE..04..010C}. Notable exceptions arise in confining systems, where quasi-particles experience an effective linear potential, suppressing their ballistic propagation and consequently inhibiting light-cone spreading~\cite{Kormosetal2017NatPh..13..246K,Fronts2024ScPP...16..138K}.

Integrable models provide a setting in which this quasi-particle picture can be made analytically precise. In particular, the quench problem admits an analytic treatment when the initial state is chosen to be an integrable boundary state~\cite{sineGquench2014JSMTE..10..035B,Integrablequench2017NuPhB.925..362P}.

In this work, we propose a boundary quench protocol that differs from those commonly considered in the literature.
We begin with the pre-quench state, defined as the ground state of a system with fixed boundary conditions. At time 
$t=0$, one of the boundary conditions is suddenly changed through the insertion of a boundary-changing operator. 
This produces a state in the Hilbert space associated with the modified boundary condition, which subsequently 
evolves under the corresponding post-quench Hamiltonian. We then investigate vacuum-to-vacuum matrix elements of 
local operators inserted after the quench. 

Spelling out the details, the object of interest is the correlation function:
\begin{equation}\label{eq:Gdef1}\nonumber
    \  _{\gamma,\beta}\bra{0} \Phi(x,t) \psi_{\beta\alpha}(0,0) \ket{0}_{\gamma,\alpha}\,,
\end{equation}
where the theory is defined on a strip. The system is initially prepared in the ground state with boundary conditions $\gamma$ and 
$\alpha$ at the left and right ends of the strip, respectively. The boundary quench is implemented at $t=0$ by the instantaneous insertion of the boundary-changing operator $\psi_{\beta,\alpha}$  at the right boundary, which switches the boundary condition from 
$\alpha$ to $\beta$.  For $t>0$, a local bulk field $\Phi(x,t)$ is inserted, and the resulting state is projected onto the ground state corresponding to boundary conditions $\gamma$ and $\beta$. 

We begin our analysis in the conformal setting, where the quench involves switching between conformally invariant boundary conditions. In this case, the vacuum-to-vacuum amplitude can be mapped to chiral correlation functions, which can be computed exactly by exploiting conformal symmetry.

We then turn to the massive integrable deformation of the conformal field theory. In this framework, we introduce an integrable relevant bulk perturbation and consider quenches between integrable boundary conditions. The post-quench dynamics is analyzed using a double form-factor expansion involving both boundary and bulk form factors. In particular, we construct matrix elements of bulk fields between boundary multiparticle states and make systematic use of form factors of boundary-changing operators.

In both the conformal and massive cases, we use the Lee–Yang model as an illustrative example, for which all necessary input data—such as conformal data and bulk and boundary form factors—are available \cite{Zamolodchikov:1990bk,Dorey:1997yg,Dorey:2000eh,Bajnok:2015iwa}.

Furthermore, in the massive theory we develop a Hamiltonian truncation approach to independently test the bulk form-factor expansion and to numerically simulate the real-time evolution following the quench.

The paper is organized as follows. In Section~\ref{sec:CFT} we analyze the boundary quench protocol in the conformal field theory. Section~\ref{sec:massive} presents the form-factor expansion for the massive integrable model. In Section~\ref{sec:TCSA} we introduce the Hamiltonian truncation method and apply it to verify the form-factor results and the time evolution. Finally, we summarize our conclusions in Section~\ref{sec:conc}. Technical details of the Hamiltonian truncation are deferred to the Appendix.

\section{Boundary quench in the CFT}\label{sec:CFT}

We first investigate the effect of quenching the boundary condition
in a boundary conformal field theory (BCFT).  We work on the strip of width $L$, which we can also take to infinity if needed.
One of the conformal boundaries is located at $x=0$, while the other
is located at $x=-L$. We insert a boundary changing operator $\psi_{\beta\alpha}$
on the $x=0$ boundary at $t=0$, which changes the boundary condition
from $\alpha$ to $\beta$. In order to study the propagation of its
effect we insert a bulk operator in a generic position $\Phi(x,t)$,
but we assume that $t>0$, as we would like to see what happens \emph{after}
we change the boundary condition.
In general, one can study the following correlator:
\begin{equation}
    G_{\gamma,\beta\alpha}^{\Phi}(x,t)=\,_{\gamma\beta}\langle 0|\Phi(x,t)\psi_{\beta\alpha}(0)|0\rangle_{\gamma\alpha}\,.
\end{equation}
or, alternatively written as
\begin{equation}\label{eq:GCFT}
    G_{\gamma,\beta\alpha}^{\Phi}(x,t)=\langle\psi_{\beta\gamma}\vert\Phi(x,t)\psi_{\beta\alpha}(0)\psi_{\alpha\gamma}(-\infty)|0\rangle\,.
\end{equation}
However, we restrict ourselves to situations where the initial or the final right boundary is chosen to be the same as the left one. In fact, this greatly simplifies the computations, since one of the boundary changing operators in~\eqref{eq:GCFT} is the identity, and the computation reduces to the determination of chiral four-point functions.

In particular, we take the left boundary to be either \emph{i)} $\gamma=\alpha$ or \emph{ii)} $\gamma=\beta$.  While these two choices are related by time reversal $t\to -t$, this transformation sends the bulk field to $\Phi(x,-t)$, which corresponds to a distinct physical setup. For this reason, we present the details separately for the two cases.
\begin{center}
\begin{figure}
\begin{centering}
\includegraphics[width=5cm]{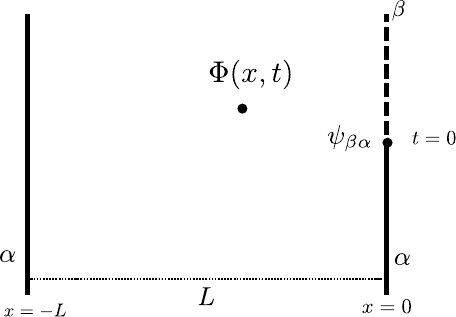}\hspace{2cm}\includegraphics[width=5cm]{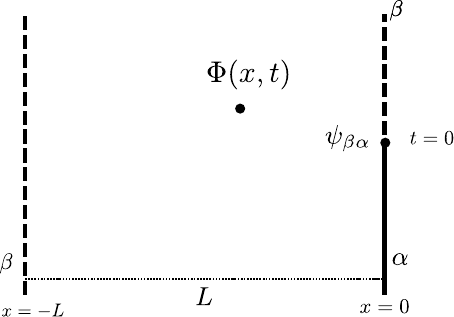}
\par\end{centering}
\caption{We work on the strip of width $L$, boundary condition $\alpha$ or
$\beta$ on the left and change the right boundary condition from
$\alpha$ to $\beta$ by inserting $\psi_{\beta\alpha}$ at $t=0$,
and monitor the effect of the change in the bulk by inserting the
operator $\Phi(x,t)$ after the change. \label{fig:BCFT1}}
\end{figure}
\par\end{center}

In the first case the left boundary is taken to be $\alpha$, which
is illustrated on the left of Figure \ref{fig:BCFT1}. We need to
calculate the following BCFT correlation function 

\begin{equation}
    G_{\alpha,\beta\alpha}^{\Phi}(x,t)=\,_{\alpha\beta}\langle 0|\Phi(x,t)\psi_{\beta\alpha}(0)|0\rangle_{\alpha\alpha}\,.
\end{equation}
where time ordering is already implemented, we change the boundary
condition first and then act with the bulk operator. 

\begin{figure}

\includegraphics[width=6cm]{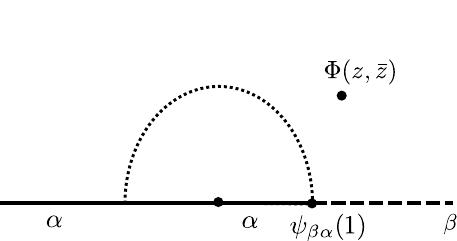}\hspace{2cm}\includegraphics[width=6cm]{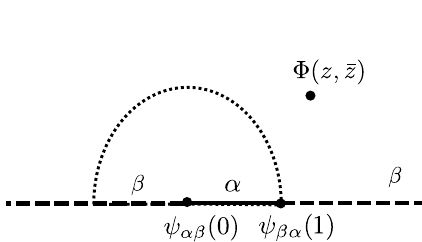}

\caption{By mapping the strip to the upper half plane we arrive at the arrangements
of operators. \label{fig:UHP1}}
\end{figure}

We investigate both the Euclidean and the Minkowskian situations and introduce light-cone and Euclidean coordinates as 
\begin{equation}
\zeta=x-t=x+iy\quad;\quad\bar{\zeta}=x+t=x-iy
\end{equation}
BCFTs are simpler on the upper half plane (UHP), so we map the strip to the UHP by the exponential map
\begin{equation}
z=e^{-i\frac{\pi}{L}\zeta}=e^{\frac{\pi}{L}(y-ix)}=u+iv\quad;\qquad\bar{z}=e^{i\frac{\pi}{L}\bar{\zeta}}=e^{\frac{\pi}{L}(y+ix)}=u-iv
\end{equation}
Under this map the remote past $y\to-\infty$ is mapped to the origin, where the left and right boundaries actually meet. Similarly, the remote future $y\to\infty$ is mapped to infinity, where the two boundaries meet again. The instant $t=y=0$ line is mapped to the upper unit circle. The boundary changing operator is located at $z=1$ as shown on the left of Figure~\ref{fig:UHP1}. 

Assuming conformal invariant boundaries, the holomorphic and anti-holomorphic
components of the energy momentum tensor coincides $T(z)\vert_{v=0}=\bar{T}(\bar{z})\vert_{v=0}$
at the boundaries. This implies that there is only one copy of the
Virasoro algebra acting on the boundary Hilbert space  and we can calculate 
all correlation functions in a chiral theory, preferably on the upper half plane. The correlation functions on the strip can be written in terms of those of the plane
by using the transformation rules of the primary fields
\begin{equation}
\Phi(\zeta,\bar{\zeta})=\left(\frac{dz}{d\zeta}\right)^{h}\left(\frac{d\bar{z}}{d\bar{\zeta}}\right)^{\bar{h}}\Phi(z,\bar{z})=(z\bar{z})^{h}\left(\frac{\pi}{L}\right)^{2h}\Phi(z,\bar{z})
\end{equation}
where we assumed that $\bar{h}=h$. The boundary field transforms
as 
\begin{equation}
\psi_{\beta\alpha}(\zeta)=\left(\frac{dz}{d\zeta}\right)^{h}\psi_{\beta\alpha}(z)=(z)^{h}\left(\frac{\pi}{iL}\right)^{h}\psi_{\beta\alpha}(z)\,.
\end{equation}
Finally, the strip correlation function is related to the one on the plane as 
\begin{equation}\label{eq:UHP1}
    G_{\alpha,\beta\alpha}^{\Phi}(x,t) \sim \left(\frac{\pi}{L}\right)^{3h}(z\bar{z})^{h}\langle\psi_{\beta\alpha}|\Phi(z,\bar{z})\psi_{\beta\alpha}(1)|0\rangle\,
\end{equation}
where $\vert 0\rangle $ is the $SL(2)$ invariant state for the chiral algebra and
\begin{equation}
    \bra{\psi_{\beta\alpha}} = \left[ \ket{\psi_{\beta\alpha}}\right]^\dagger = \lim_{w\rightarrow\infty} w^{2h}\bra{0}\psi_{\alpha\beta}(w) \
\end{equation}
implements the boundary changing at infinity. Correlation functions on the UHP are understood in radial ordering, which is the consequence of the time ordering on the strip. We assume that $\vert z\vert>1$ as we would like to see how the expectation value changes after the boundary quench. 

The situation, when the left boundary is $\beta$ is slightly different.
In this case there is no boundary changing operator at future infinity, as the boundary conditions agree there. 
In contrast, at past infinity the $\beta$ and $\alpha$ boundaries meet so we
need to insert a non-trivial operator. These differences are demonstrated on the right of Figure \ref{fig:BCFT1} and \ref{fig:UHP1}. The strip correlation
function is related to the one on the plane as 
\begin{equation}\label{eq:UHP2}
    G_{\beta,\beta\alpha}^{\Phi}(x,t) \sim \left(\frac{\pi}{L}\right)^{3h}(z\bar{z})^{h}\langle 0|\Phi(z,\bar{z})\psi_{\beta\alpha}(1)\psi_{\alpha\beta}(0)|0\rangle\,
\end{equation}
In the following we determine these correlators in the Lee-Yang model
from conformal invariance. 

\subsection{Lee--Yang CFT: a quick recap}

To set the stage for later computations, we quickly review the Lee--Yang boundary CFT \cite{Dorey:1997yg, Dorey:2000eh,Bajnok:2020kyk}.

The critical Lee--Yang model is the non-unitary minimal model $\mathcal{M}_{2,5}$ with central charge $c=-\frac{22}{5}$. It has two Virasoro representations of Kac labels $(1,1)$ and $(2,1)$. The first one $V_\mathbb{I}$ is related to the identity $\mathbb{I}$, while the second, $V_\varphi$ is related to the primary field $\varphi$ with conformal weight $h=-\frac{1}{5}$.

 Conformal covariance restricts the chiral four-point function, of four insertions of $\varphi$,  to the form 
\begin{equation}
\langle0\vert\varphi(z_{1})\varphi(z_{2})\varphi(z_{3})\varphi(z_{4})\vert0\rangle=\prod_{1\leq i<j}z_{ij}^{-2h+\frac{4h}{3}}f(x)\quad;\quad x=\frac{z_{12}z_{34}}{z_{13}z_{24}}\,.
\end{equation}
From the definition of the out state $\langle\varphi\vert=\lim_{z_{1}\to\infty}z_{1}^{2h}\langle0\vert\varphi(z_{1})$
we can simplify our expression as
\begin{equation}
\langle\varphi\vert\varphi(z)\varphi(\bar{z})\varphi(1)\vert0\rangle=((z-\bar{z})(z-1)(\bar{z}-1))^{-\frac{2h}{3}}f\Bigl(\frac{\bar{z}-1}{z-1}\Bigr)\,.
\end{equation}
To determine $f(z)$, we recall the usual representation
of the chiral four-point function
\begin{equation}\label{eq:4ptsimple}
\langle\varphi\vert\varphi(1)\varphi(z)\vert\varphi\rangle=((1-z)z)^{-\frac{2h}{3}}f(z)\,.
\end{equation}
and the fact that $\varphi$ is degenerate at level $2$, which implies
\begin{equation}
\langle\varphi\vert\varphi(1)\varphi(z)\left[L_{-1}^{2}-\frac{2}{5}L_{-2}\right]\vert\varphi\rangle=0\,,
\end{equation}
This leads to a second-order differential equation, see e.g. in \cite{Dorey:2000eh,Bajnok:2013waa}, which has the generic solution
\begin{equation}\label{eq:chir4pt}
\langle\varphi\vert\varphi(1)\varphi(z)\vert\varphi\rangle=c_{1}(z(1-z))^{-h}F_{2,1}\Bigl(\frac{1}{5},\frac{2}{5};\frac{4}{5}\Big \vert z\Bigr)+c_{2}(z^{2}(1-z))^{-h}F_{2,1}\Bigl(\frac{2}{5},\frac{3}{5};\frac{6}{5} \Big \vert z\Bigr)\,.
\end{equation}
Combining~\eqref{eq:chir4pt} with~\eqref{eq:4ptsimple} provides us with $f$. 

Corresponding to the two heighest weight representations of the Virasoro algebra, there are two conformally invariant boundary conditions in the Lee--Yang theory, which we denote by $\mathbb{I}$ and $\phi$. Consequently, on a strip geometry we have three possible arrangements: $\mathbb{I}\mathbb{I},\mathbb{I}\phi$ and $\phi \phi$, with the corresponding boundary Hilbert spaces $\mathcal{H}_{\mathbb{I}\mathbb{I}}=V_0,\mathcal{H}_{\mathbb{I}\phi}=V_{\varphi}$ and $\mathcal{H}_{\phi,\phi}=V_0\oplus V_\varphi$. 

For the $\mathbb{I}$ boundary only fields corresponding to the vacuum representation live, while for the $\phi$ boundary, we have both the vacuum and the $h$ representation. The corresponding primary field, which lives on the boundary is denoted by $\varphi$ and the bulk-boundary operator product expansion (OPE) takes the general form
\begin{equation}\label{eq:buboOPE}
\Phi(z,\bar{z})=(2v)^{-2h}B_{\gamma,\mathbb{I}}+(2v)^{-h}B_{\gamma,\varphi}\varphi(u)+\dots\,,
\end{equation}
where we introduced the notation $z,\bar z=u\pm i v$ and the ellipses represent descendant fields and $\gamma$ labels the boundary conditions. This also implies the short-distance behaviour
\begin{equation}
\langle\psi_{\beta\alpha}\vert\Phi(z,\bar{z})\psi_{\beta\alpha}(1)\vert0\rangle\sim(2v)^{-2h}B_{\gamma,\mathbb{I}}+(2v)^{-h}B_{\gamma,\varphi}\langle\psi_{\beta\alpha}\vert\varphi(u)\psi_{\beta\alpha}(1)\vert0\rangle+\dots\,,
\end{equation}
where for $u<1$ the boundary condition is $\gamma=\alpha$, while for $u>1$ it is $\gamma=\beta$. The conformal property of the boundary 3-point function dictates that
\begin{equation}
\langle\psi_{\beta\alpha}\vert\varphi(u)\psi_{\beta\alpha}(1)\vert0\rangle=\lim_{w\to\infty}w^{2h}\langle0\vert\psi_{\alpha\beta}(w)\varphi(u)\psi_{\beta\alpha}(1)\vert0\rangle=C_{{\psi}_{\alpha\beta}\varphi\psi_{\beta\alpha}}(u-1)^{-h}
\end{equation}
and we assumed canonical normalization $\langle\psi_{\beta\alpha}\vert\psi_{\beta\alpha}\rangle=1$. 

There is no non-trivial boundary field living on the $\mathbb I$ boundary, therefore the bulk-boundary coupling $B_{\mathbb{I},\varphi}$ is $0$. The non-zero couplings are given as~\cite{Dorey:1997yg,Dorey:2000eh}
\begin{equation}\label{eq:bubo}
B_{\mathbb{I},\mathbb{I}}=-\left(\frac{2}{1+\sqrt{5}}\right)^{\frac{1}{2}}\quad;\quad B_{\phi,\mathbb{I}}=\left(\frac{2}{1+\sqrt{5}}\right)^{-\frac{3}{2}}\quad;\quad B_{\phi,\varphi}=\left(\frac{2}{5+\sqrt{5}}\right)^{-\frac{1}{2}}\rho\,,
\end{equation}
and
\begin{equation}\label{eq:bobo}
C_{{\psi}_{\beta\alpha}\varphi\psi_{\beta\alpha}}=-\left(\frac{2}{1+\sqrt{5}}\right)^{\frac{1}{2}}\rho \quad;\quad\rho=\left(\frac{\Gamma(\frac{1}{5})\Gamma(\frac{6}{5})}{\Gamma(\frac{3}{5})\Gamma(\frac{4}{5})}\right)^{\frac{1}{2}}\,.
\end{equation}

Let us emphasise a particular feature of the Lee–Yang model. Because $h=-1/5$, the short-distance behaviour of the expectation value remains bounded, whereas at large distances it actually increases. This mirrors the behaviour observed in the bulk two-point function. It also suggests that the genuinely physical observables are better thought of as derivatives of the fields, rather than the fields themselves. This is reminiscent of the free scalar field theory, where the two-point function grows logarithmically, while its derivatives decay. 

\subsection{Correlation functions on the plane}

We now use conformal invariance to determine the space-time dependence
of the correlators. First, we demonstrate the method starting with the first situation with $\gamma=\alpha$, i.e. changing from the same boundary conditions to different ones.

Since in the boundary theory we have only one copy of the Virasoro algebra, the correlation function (\ref{eq:UHP1}) transforms the same way as a chiral 4-point function \cite{Cardy:2004hm}:
\begin{equation}
\langle\psi_{\beta\alpha}|\Phi(z,\bar{z})\psi_{\beta\alpha}(1)|0\rangle\sim\langle\varphi\vert\varphi(z)\varphi(\bar{z})\varphi(1)\vert0\rangle\,,
\end{equation}
where all fields are chiral primaries of dimension $h$, but one of
the insertion points is $\bar{z}$, which lives in the lower half plane.
The correlation function relevant for the boundary problem takes the form 
\begin{align}
\nonumber \langle\varphi\vert\varphi(z)\varphi(\bar{z})\varphi(1)\vert0\rangle & =c_{1}(z-\bar{z})^{\frac{1}{5}}(\bar{z}-1)^{\frac{1}{5}}F_{2,1}\Bigl(\frac{1}{5},\frac{2}{5};\frac{4}{5}\Big \vert\frac{\bar{z}-1}{z-1}\Bigr)\\
 & \quad+c_{2}\frac{(z-\bar{z})^{\frac{1}{5}}(\bar{z}-1)^{\frac{2}{5}}}{(z-1)^{\frac{1}{5}}}F_{2,1}\Bigl(\frac{2}{5},\frac{3}{5};\frac{6}{5}\Big \vert\frac{\bar{z}-1}{z-1}\Bigr)
\end{align}
where we choose a specific cut structure in order to have smooth behaviours
away from the boundaries. We need to find the right linear combination,
which corresponds to changing the boundary at $z=1$ in the required way. In doing so we take $z=u+iv$ and investigate the small $v$ behaviour.  When $v\to0$ the bulk operator approaches the boundary and the bulk-boundary OPE~\eqref{eq:buboOPE} can be used. It should be different for $u\lessgtr1$ as they correspond to different boundary conditions.

In the situation when we switch from the $\alpha=\mathbb{I}$ boundary
to the $\beta=\phi$ boundary we demand that for $u<1$ the correlation
function is compatible with the vanishing OPE coefficient $B_{\mathbb{I},\varphi}=0$.
This implies that $c_{1}$ and $c_{2}$ are not independent. In particular, with the proper normalization we found that the choice
\begin{equation}\label{eq:CCs1}
c_{1}=-\frac{(-2)^{\frac{1}{5}}5^{\frac{7}{4}}\Gamma(\frac{1}{5})\Gamma(\frac{11}{10})\Gamma(\frac{6}{5})}{(5+\sqrt{5})\pi^{\frac{3}{2}}}\quad;\quad c_{2}=\frac{(-1)^{\frac{3}{5}}2^{\frac{1}{5}}5^{\frac{3}{4}}\sqrt{\pi}\Gamma(\frac{1}{10})\Gamma(\frac{6}{5})}{\Gamma(\frac{1}{5})^{2}\Gamma(\frac{2}{5})}\,,
\end{equation}
reproduces both~\eqref{eq:bubo} and~\eqref{eq:bobo}.

\begin{figure}
\includegraphics[width=0.45\textwidth]{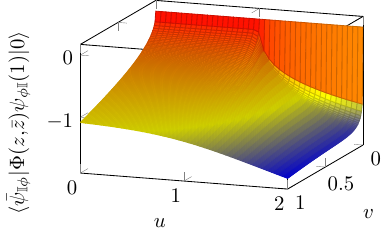}\hspace{1cm}\includegraphics[width=0.45\textwidth]{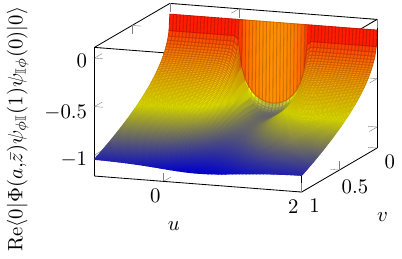}
\caption{3D plot of the correlation functions $\langle\psi_{\mathbb{I}\phi}\vert\Phi(z,\bar{z})\psi_{\phi\mathbb{I}}(1)\vert0\rangle$
(left) and $\langle0\vert\Phi(z,\bar{z})\psi_{\phi\mathbb{I}}(1)\psi_{\mathbb{I}\phi}(0)\vert0\rangle$
(on the right). We can recognize the change in the boundary condition at $z=1$
and at $z=0,1$ respectively. \label{fig:Eucplane}}
\end{figure}

We visualize the above correlation function in the Euclidean case on the plane by the 3D plot in the left panel of Figure \ref{fig:Eucplane}. This clearly indicates the change of the exponent, after changing the boundary condition.

We can now focus on the situation when we start with different boundaries and change the right boundary to the same as the left, i.e. $\gamma=\beta$. This situation is described by the previously introduced function $f(z)$ as:
\begin{equation}
    \langle0|\psi_{\beta\alpha}(1)\Phi(z,\bar{z})\psi_{\alpha\beta}(0)|0\rangle\sim((1-z)(1-\bar{z})(z-\bar{z})z\bar{z})^{-\frac{2h}{3}}f\left(\frac{(1-z)\bar{z}}{(1-\bar{z})z}\right)\,.
\end{equation}
We now choose $\beta=\mathbb{I}$ and $\alpha=\phi$, so that we change the boundary condition
from $\phi$ to $\mathbb{I}$. To fix the correct behaviour at various segments of the boundary the coefficients in $f$ are chosen to be
\begin{equation}
c_{2}=\frac{\sqrt{\frac{6}{\sqrt{5}}-2}\pi}{\Gamma(\frac{1}{5})\Gamma(\frac{4}{5})-\Gamma(\frac{2}{5})\Gamma(\frac{3}{5})}\quad;\quad c_{1}=-\frac{\Gamma(\frac{2}{5})\Gamma(\frac{6}{5})}{\Gamma(\frac{4}{5})^{2}}c_{2}\,.
\end{equation}
The physical correlation function $\langle0|\Phi(z,\bar{z})\psi_{\mathbb{I}\phi}(1)\psi_{\phi\mathbb{I}}(0)|0\rangle$ is the analytic continuation in $z$ and $\bar z$  of the function we just determined $\langle0|\psi_{\mathbb{I}\phi}(1)\Phi(z,\bar{z})\psi_{\phi\mathbb{I}}(0)|0\rangle$ and has the correct boundary limits. For $0<u<1$ the boundary condition is $\phi$ so we have an exponent corresponding to the boundary field $\varphi$ and $B_{\phi,\varphi}\neq0$, while for $u<0$ and $u>1$ the corresponding term vanishes $B_{\mathbb{I},\varphi}=0$. This behaviour is clearly visible on the right panel of Figure \ref{fig:Eucplane}. 
Let us now examine how the physically relevant correlation functions behave on the strip.

\subsection{Correlation functions on the strip}

We return to our original setup and transform the correlators from the plane back to the strip:
\[
  G_{\alpha,\beta\alpha}^{\Phi}(x,t)\sim\left(\frac{\pi}{L}\right)^{3h}(z\bar{z})^{h}\langle\psi_{\beta\alpha}\vert\Phi(z,\bar{z})\psi_{\beta\alpha}(1)\vert0\rangle
\]Let us recall that the strip correlators in cases \emph{i)} and \emph{ii)} are connected to the UHP correlators by eqs.~\eqref{eq:UHP1} and~\eqref{eq:UHP2}, respectively. Specifically, up to an overall phase, they can be written in terms of the functions derived in the previous subsection by making the appropriate substitutions for $z$ and $\bar z$.

In the Euclidean setting, $z=e^{\frac{\pi}{L}(y-ix)}$ and $\bar{z}=e^{\frac{\pi}{L}(y+ix)}$, the behaviour is qualitatively the same as on the plane if we focus only on the boundary where the boundary change happens. 

\begin{figure}

\includegraphics[width=0.45\textwidth]{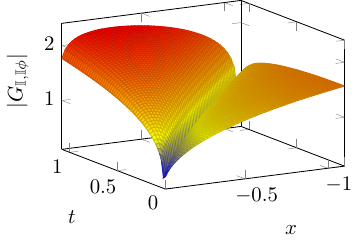}\hspace{1cm}\includegraphics[width=0.45\textwidth]{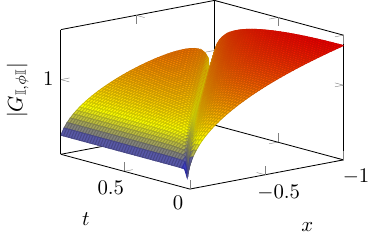}
\caption{The correlator of the bulk field after the boundary quench in the
Minkowski strip. The volume is $L=10$ and we focused on the $-1\protect\leq x\protect\leq0$
and $0<t<1$ domain. Quenching from $\mathbb{II}$ to $\mathbb{I}\phi$ is on the left, while from $\mathbb{I\phi}$ to $\mathbb{II}$ is
on the right. \label{fig:MinCFTquench}}

\end{figure}

However, in the Minkowski setting ($z=e^{i\frac{\pi}{L}(t-x)}$ , $\bar{z}=e^{i\frac{\pi}{L}(t+x)}$) the situation is more interesting. We present the results in Figure \ref{fig:MinCFTquench} for the two quenches considered. The volume is taken to be relatively large $L=10$, in order to avoid finite size effect and we focus on the space domain $-1\leq x\leq0$ and the time interval $0<t<1$ after the boundary quench. We plot the modulus of the expectation value of the bulk field. Clearly the effect of the boundary quench propagates with the speed of light and we can see the light-cone structure.

\begin{figure}
    \centering
    \includegraphics{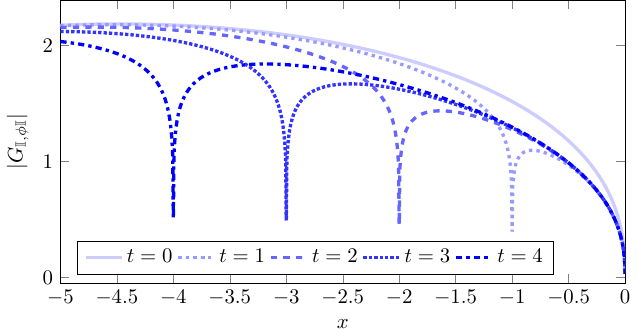}
    \caption{Space dependence of the amplitude for different times for the quench $\mathbb I\phi \rightarrow \mathbb{II}$ in volume $L=10$. The light-cone singularity is visible for $x=-t$, as expected.}
    \label{fig:timeslices}
\end{figure}

The light-cone effect is also apparent in Figure~\ref{fig:timeslices}, where we show several time slices for the quench $\mathbb I\phi \rightarrow \mathbb{II}$ in a system of size $L=10$. On the right-hand side of the figure, close to the boundary, the modification of the bulk–boundary OPE is clearly visible: the curves at $t\neq0$ approach the boundary with different power laws.

\begin{figure}
    \centering
    \includegraphics{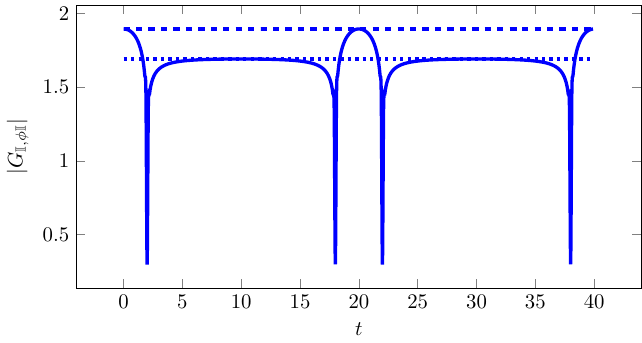}
    \caption{Time dependence of the amplitude for $x=-2$ for the quench $\mathbb I\phi \rightarrow \mathbb{II}$ in volume $L=10$. We indicate the VEV at $t=0$ (dashed) and $t=10$ (dotted). The light-cone bounces back and forth between the two boundaries, periodically saturating to the two values, depending on the relative space-time position of the light-cone.}
    \label{fig:spaceslice}
\end{figure}

Similarly, we plot the time dependence at $x=-2$ for demonstration in Figure~\ref{fig:spaceslice}, for somewhat larger times. The value of the VEV starts with the one with the initial boundary conditions, then at $t=|x|$ the light-cone hits and the VEV saturates to the another value. At $t=L$ the light-cone bounces back from the left boundary, and hits the specific point again, and the VEV saturates to its initial value, until the light-cone gets reflected again from the right boundary. This pattern continues: the light-cone bounces back and forth between the two boundaries. This is a purely Minkowski effect, due to the time-periodicity of our formulas evaluated in Minkowski time. Similar revival effect was also observed in other quenched CFT problems \cite{2014PhRvL.112v0401C,Janik:2025zji}. Contrary, in the Euclidean setting, the factor $(z\bar z)^h$ leads to exponential damping in the Euclidean time, since $h<0$.

\section{Boundary quench in the massive theory}\label{sec:massive}

In this section, we investigate the boundary quench in an integrable massive deformation of a boundary conformal field theory. We perform our analysis on the semi-infinite plane rather than on a strip geometry, as our focus is on the intrinsic quench dynamics and not on boundary-induced finite-size effects. The boundary is located at $x=0$, and the action of the boundary scaling Lee–Yang model is given by
\begin{equation}
S = S_{\mathrm{BCFT}} + \lambda \int dt \int_{-\infty}^{0} dx\, \Phi(x,t) + \lambda_b \int dt\, \varphi(t)\,,
\end{equation}
where $\varphi(t)$ denotes a boundary primary field, which exists only for the $\phi$-type boundary condition.
In what follows, we set $\lambda_b=0$, i.e., we do not include a boundary perturbation. From the perspective of integrability, however, extending the analysis to $\lambda_b\neq0$ is straightforward. This choice allows us to treat in a unified manner the integrable bulk perturbations of the BCFT with boundary conditions $\mathbb{I}$ and $\phi $. We adopt the same labeling of boundary conditions in the massive theory as in the underlying BCFT.

The resulting (bulk) theory is a paradigmatic example of integrable quantum field theories, where the scattering is purely elastic and factorized to two-particle processes.  The parameter $\lambda$ is dimensionful, which sets the mass scale in the perturbed theory~\cite{1995IJMPA..10.1125Z} 
\begin{equation}
m=\kappa\lambda^{\frac{1}{2-2h}}\quad;\quad\kappa=2^{\frac{19}{12}}\sqrt{\pi}\frac{\left(\Gamma(\frac{3}{5})\Gamma(\frac{4}{5})\right)^{\frac{5}{12}}}{5^{\frac{5}{16}}\Gamma(\frac{2}{3})\Gamma(\frac{5}{6})}=2.642944\label{eq:kappa}\,.
\end{equation}
The Lee--Yang scattering theory has only a single type of particle which scatter on itself with the scattering phase~\cite{1989PhLB..225..275C} 
\begin{equation}
S(\theta)=\frac{\sinh\theta+i\sin\frac{\pi}{3}}{\sinh\theta-i\sin\frac{\pi}{3}}=-\left(\frac{1}{3}\right)\left(\frac{2}{3}\right)\,,
\end{equation}
where we also introduced the block notation 
\begin{equation}
(x)=\frac{\sinh(\frac{\theta}{2}+i\frac{\pi x}{2})}{\sinh(\frac{\theta}{2}-i\frac{\pi x}{2})}\,.
\end{equation}

In the presence of a boundary, particles reflect off the boundary. Notably, the presence of the boundary preserves integrability \cite{Ghoshal:1993tm},  leading to similar properties of the boundary reflection as of the S-matrix. In particular, the boundary reflection amplitude for the $\mathbb{I}$ boundary is~\cite{Dorey:1997yg}
\begin{equation}
R_{\mathbb{I}}(\theta)=\left(\frac{1}{2}\right)\left(\frac{1}{6}\right)\left(-\frac{2}{3}\right)\,,
\end{equation}
which exhibits a pole at $i\frac{\pi}{2}$ with the corresponding boundary coupling $g_{\mathbb{I}}=-2i\sqrt{(2\sqrt{3}-3)}$. This shows that the identity boundary $\mathbb{I}$ can emit a virtual particle with zero energy.

The reflection factor of the $\phi$ boundary depends on  the boundary parameter $b$
 as \begin{equation}
R_{\phi}(\theta)=R_{\mathbb{I}}(\theta)R_{b}(\theta)\quad,\quad R_{b}(\theta)=S(\theta-\theta_{0})S(\theta+\theta_{0})\quad,\quad\theta_{0}=i\pi\frac{3-b}{6},
\end{equation}
which is related to the boundary perturbation as 
\begin{equation}
\lambda_b=-h_{{\rm crit}}\sin(\pi(b+1/2)/5)m^{6/5}\,.
\end{equation}
 where $h_{\mathrm{crit}}=0.68529$. 
The $\phi$ boundary can also emit a virtual zero energy particle with amplitude 
\begin{equation}
g_{\phi}\left(b\right)=\frac{\cosh\theta_{0}+\sin\frac{\pi}{3}}{\cosh\theta_{0}-\sin\frac{\pi}{3}}g_{\mathbb{I}} \  .
\end{equation}
Note that $R^{\mathbb{I}}(\theta)$ is identical to $R^{\phi}(\theta)$
at $b=0$ and so both have a pole at $\theta=\frac{i\pi}{2}$ coming
from the $\left(\frac{1}{2}\right)$ block, but their $g$ factors
differ in sign. The value, when there is no boundary perturbation
in the action, corresponds in the scattering theory to $b=-\frac{1}{2}$, so this is the case we restrict our analysis. 

So far the boundary was localized in space, i.e. at $x=0$ for all $t$.
Alternatively, one can make a double Wick rotation and place the boundary
at a given time slice. In this setting the boundary emits particles
via a boundary state. This state is an element of the bulk Hilbert space and can be calculated from the reflection
factors as \cite{Ghoshal:1993tm,Dorey:1997yg,Bajnok:2006dn}
\begin{equation}
\vert B\rangle_{\alpha}=\exp\left\{ \frac{g_{\alpha}}{2}A^{+}(0)+\int_{0}^{\infty}\frac{d\vartheta}{2\pi} \ R_{\alpha}(\frac{i\pi}{2}+\vartheta)A^{+}(\vartheta)A^{+}(-\vartheta)\right\} \vert0\rangle\label{eq:Bstate}
\end{equation}
where $A^{+}(\vartheta)$ emits a particle with rapidity $\vartheta$
in the Wick rotated theory. By using the $ R^{\alpha}(i\pi/2-\theta)=S(2\theta) R^{\alpha}(i\pi/2+\theta)$ boundary crossing symmetry of the reflection factor one can show that the integrand is a symmetric function of $\vartheta$ and the integral domain can be extended for the whole line. 

\subsection{Time evolution after the boundary quench}

\begin{figure}
\begin{centering}
\includegraphics[width=5cm]{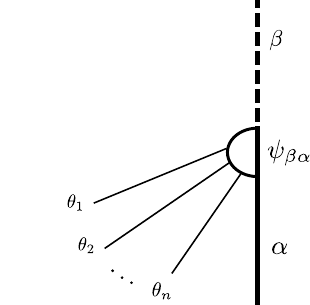}\hspace{2cm}\includegraphics[width=5cm]{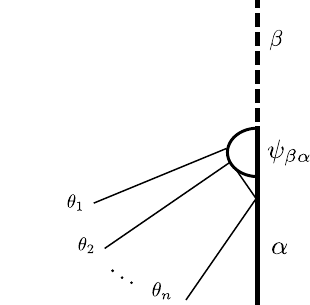}
\par\end{centering}
\caption{Graphical representation of the form factor of a boundary changing
operator (half bulb) and its reflection property. \label{fig:bff} }

\end{figure}

In the following we investigate the boundary quench in the massive
integrable field theory. In particular, we would like to calculate
the overlap of the eigenstates of the post quench Hamiltonian with
the pre quench groundstate. These observables are related to the matrix
elements of the boundary changing operator between asymptotic multiparticle
states: 
\begin{equation}
_{\beta}\langle\theta_{1},\dots,\theta_{n}\vert\psi_{\beta\alpha}(t)\vert0\rangle_{\alpha}=e^{i(\Delta E_{\beta\alpha}+\sum_{j}m\cosh\theta_{j})t}{}_{\beta}\langle\theta_{1},\dots,\theta_{n}\vert\psi_{\beta\alpha}\vert0\rangle_{\alpha}
\end{equation}
where we abbreviated $\psi_{\beta\alpha}(0)$ as $\psi_{\beta\alpha}$. These
matrix elements are called boundary form factors, which all can be expressed
in terms of the elementary ones \cite{Bajnok:2015iwa}
\begin{equation}
_{\beta}\langle\theta_{1},\dots,\theta_{n}\vert\psi_{\beta \alpha}\vert0\rangle_{\alpha}=F_{n}^{\mathcal{\psi}_{\beta \alpha}}(\theta_{1}+i\pi ,\theta_{2}+i\pi ,\dots,\theta_{n}+i\pi )\,.
\end{equation}
The elementary form factors contain incoming states rather than outgoing ones and can be visualized
on Figure \ref{fig:bff}. We also presented graphically the reflection
axiom of the elementary form factors 
\begin{equation}
F_{n}^{\psi_{\beta \alpha}}(\theta_{1},\theta_{2},\dots,\theta_{n})\equiv  {}_{\beta}\langle0\vert\psi^{\beta\alpha}\vert\theta_{1},\dots,\theta_{n}\rangle_{\alpha}=R_{\alpha}(\theta_{n})F_{n}^{\psi_{\beta\alpha}}(\theta_{1},\theta_{2},\dots,-\theta_{n})\,,
\end{equation}
which reflects the fact that the state $\vert\theta_1,\dots,\theta_n \rangle_\alpha$  is equivalent to 
$R_\alpha(\theta_n)\vert\theta_1,\dots, -\theta_n\rangle_\alpha$. 

The elementary form factors of the operator, which changes between boundary
conditions $\beta$ and $\alpha$ satisfies the form factor axioms of \cite{Bajnok:2015iwa}. These include the reflection axioms already mentioned, its crossed version and the permutation axiom, which reads as 
\begin{equation}
F_{n}^{\psi_{\beta \alpha}}(\theta_{1},\dots,\theta_{i},\theta_{i+1},\dots,\theta_{n})=S(\theta_{i}-\theta_{i+1})F_{n}^{\psi_{\beta\alpha}}(\theta_{1},\dots,\theta_{i+1},\theta_i,\dots,\theta_{n})\,.
\end{equation}
The exchange property indicates that the state $\vert\theta_1,\dots,\theta_n \rangle_\alpha$ is actually equivalent to any permuted variants, which further can be combined with the reflection property.

In particular, for two particles we have eight equivalent terms, since each particle can come with their reflected variants and their orders can be exchanged. Out of these eight combinations only the in state $\vert \theta_1,\theta_2 \rangle_\alpha $ and the out state $\vert -\theta_1,-\theta_2 \rangle_\alpha$ are physical and they are connected through a chain of reflection and exchange relations via the two particle reflection matrix:
\begin{align}
\vert \theta_1,\theta_2 \rangle_\alpha =&R_\alpha(\theta_2)\vert \theta_1,-\theta_2 \rangle_\alpha=S(\theta_1+\theta_2)R_\alpha(\theta_2)\vert -\theta_2,\theta_1 \rangle_\alpha
\nonumber  \\  =&
S(\theta_1+\theta_2)R_\alpha(\theta_1)R_\alpha(\theta_2)\vert -\theta_2,-\theta_1 \rangle_\alpha \nonumber  \\ =&
S(\theta_1+\theta_2)S(\theta_1-\theta_2)R_\alpha(\theta_1)R_\alpha(\theta_2)\vert -\theta_1,-\theta_2 \rangle_\alpha.
\label{eq:2ptstate}
\end{align} 

Thus a boundary state with rapidities $\theta_1,\dots,\theta_n$
contains all reflected and permuted $2^n n!$ companion  states as well. Strictly speaking,  out of these combinations only the ($\theta_i>\theta_{i+1}$) ordered in state  $\vert\theta_1,\dots,\theta_n \rangle_\alpha$  and the  $\vert-\theta_1,\dots,-\theta_n \rangle_\alpha$ out state is physical. The permutation and reflection properties tell us how to continue analytically the form factor to unphysical arguments to reach the out state from the in state consistently with the multiparticle reflection matrix, which is a product of one-particle reflections and pairwise scatterings. 

The elementary boundary form factors, when analytically extended, are meromorphic functions of the rapidities with prescribed pole singularities coming from the boundary and bulk kinematical singularities and bound state conditions. The solutions to these axioms can be parametrized as 
\begin{equation}
F_{n}^{\psi_{\beta\alpha}}(\theta_{1},\theta_{2},\dots,\theta_{n})=\left\langle \psi_{\beta\alpha}\right\rangle H_{n}^{\beta\alpha}\prod_{i=1}^{n}\frac{r^{\beta\alpha}(\theta_{i})}{y_{i}}\prod_{i<j}\frac{f(\theta_{i}-\theta_{j})f(\theta_{i}+\theta_{j})}{(y_{i}+y_{j})}Q_{n}^{\psi_{\beta\alpha}}(y_{1},y_{2}\dots,y_{n})\,,
\end{equation}
where $\left\langle \psi_{\beta\alpha}\right\rangle $ is the vacuum matrix element of the boundary changing operator, $H_{n}^{\beta\alpha} $ are normalization constant, $ y_i=2 \cosh \theta_i$, while $f(\theta)$ and $r^{\beta\alpha}(\theta)$ are the minimal bulk and boundary form factors. The symmetric polynomials $Q_{n}^{\psi_{\beta\alpha}}(y_{1},y_{2}\dots,y_{n})$ carry the operator dependence and can be calculated from the singularity axioms, recursively. 

The minimal bulk two-particle form factor  satisfies 
\begin{equation}
f(\theta)=S(\theta) f(-\theta) \quad ;\quad 
f(i\pi -\theta )=f(i\pi +\theta)
\end{equation}
and guaranties the permutation property of the form factor. In the Lee--Yang model it can be written in the form \cite{Zamolodchikov:1990bk}
\begin{equation}
f(\theta)=\frac{y-2}{y+1}v(i\pi-\theta)v(-i\pi+\theta)\quad,\quad y=e^{\theta}+e^{-\theta}\,,
\end{equation}
 where 
\begin{equation}
v(\theta)=\exp\left\{ 2\int_{0}^{\infty}\frac{dt}{t}e^{i\frac{\theta t}{\pi}}\frac{\sinh\frac{t}{2}\sinh\frac{t}{3}\sinh\frac{t}{6}}{\sinh^{2}t}\right\}\,.
\end{equation}
The minimal one-particle boundary changing form factor satisfies 
\begin{equation}
r^{\beta\alpha}(\theta)=R_{\alpha}(\theta) r^{\beta\alpha}(-\theta)\quad ;\quad r^{\beta\alpha}(i\pi +\theta)=R_\beta (-\theta) r^{\beta\alpha}(i\pi -\theta)\,.
\end{equation}

In the Lee--Yang model there is a single non-trivial boundary changing, and the corresponding minimal form factor can be written as \cite{Bajnok:2015iwa}
\begin{equation}
r^{\mathbb{I}\phi}(\theta)=r^{\mathbb{II}}(\theta)r_{\phi}(\theta)\label{eq:r_id_phi_param}\,,
\end{equation}
with 
\begin{equation}
r^{\mathbb{II}}(\theta)=4i\sinh\theta\exp\left\{ \int_{0}^{\infty}\frac{dt}{t}\frac{\sinh(t)-\cos\left(\frac{it}{2}-\frac{\theta t}{\pi}\right)\left(\sinh\frac{5t}{6}+\sinh\frac{t}{2}-\sinh\frac{t}{3}\right)}{\sinh\frac{t}{2}\,\sinh t}\right\} 
\end{equation}
and
\begin{equation}
\hspace{-0.24cm}r_{\phi}(\theta)=\mathcal{N}\exp\Big\{2\int_{0}^{\infty}\frac{dt}{t}\frac{\sinh\frac{b+1}{6}t+\sinh\frac{b-1}{6}t-\sinh\frac{b+7}{6}t-\sinh\frac{b+5}{6}t}{\sinh^{2}t}\sin^{2}\Big(\frac{i\pi-\theta}{2\pi}t\Big)\Big\}\,.
\end{equation}
The factor $\mathcal{N}$ is given as
\begin{equation}
\mathcal{N}=-\frac{1}{4}\exp\left\{ 2\int_{0}^{\infty}\frac{dt}{t}\frac{\cosh\left(\frac{b+3}{6}t\right)\left[\sinh\frac{t}{3}+\sinh\frac{2t}{3}\right]-\sinh(t)}{\sinh^{2}(t)}\right\}\,.
\end{equation}
Moreover, the normalization factor is $H_{n}^{\psi_{\mathbb{I}\phi}}=H_{n}^{\psi_{\phi\mathbb{I}}}=H_n=\left(\frac{i\sqrt[4]{3}}{v\left(0\right)\sqrt{2}}\right)^{n}$.

The solution for the operator-dependent symmetric polynomials of the energy-momentum tensor with identity boundary condition was determined in \cite{Hollo:2014vpa} as
\begin{equation}
Q_{1}^{T}=\sigma_{1}^{(1)}\quad;\qquad Q_{2}^{T}=\sigma_{1}^{(2)}\quad;\qquad Q_{3}^{T}=\left(\sigma_{1}^{(3)}\right)^{2}\quad;\qquad Q_{n}^{T}=\left(\sigma_{1}^{(n)}\right)^{2}\det\Xi^{(n)}\,,
\end{equation}
 where for $n\geq4$ the $(n-3)\times(n-3)$ matrix $\Xi^{(n)}$ is
defined as
\begin{equation}
\Xi_{ij}^{(n)}=\sum_{k\in\mathbb{Z}}3^{k}\binom{i-j+k}{k}\sigma_{3j-2i+1-2k}^{(n)}\qquad,\qquad1\leq i,j\leq n-3\,,
\end{equation}
and $\sigma$ denotes the symmetric polynomials
\begin{equation}
\prod_{i=1}^n(y+y_i)=\sum_{k=1}^n y^k \sigma^{(n)}_{n-k}\,.
\end{equation}

The solution for the boundary  changing operators $\psi_{\mathbb{I}\phi}$ and $\psi_{\phi \mathbb{I}}$ with the appropriate
initial conditions, $Q_{0}^{\psi_{\mathbb{I}\phi}}=Q_{0}^{\psi_{\phi\mathbb{I}}}=1$,  can be obtained as  
\begin{equation}
Q_{n}^{\psi_{\mathbb{I}\phi}}(y_{1},\dots,y_{n})=\left.\frac{Q_{n+1}^{T}}{\sigma_{1}^{(n+1)}}\right|_{(y_{0},y_{1},\dots,y_{n})}\quad,\quad Q_{n}^{\psi_{\phi\mathbb{I}}}(y_{1},\dots,y_{n})=\left.\frac{Q_{n+1}^{T}}{\sigma^{(n+1)}_{1}}\right|_{(-y_{0},y_{1},\dots,y_{n})}\,.
\end{equation}

Using all these ingredients the overlap between the initial vacuum and an arbitrary outgoing state can be calculated. These building blocks then can be used to determine the time evolution after the boundary quench as 
\begin{align}
\nonumber e^{iH_{\beta}t}\psi_{\beta\alpha}\vert0\rangle_{\alpha}= & \sum_{\vert n\rangle_{\beta}\in{\cal H}_{\beta}}e^{iE_{n}t}\vert n\rangle_{\beta}\ _{\beta}\langle n\vert\psi_{\beta\alpha}\vert0\rangle_{\alpha}\\ \nonumber
= & \sum_{n=0}e^{iE_{\beta}t}\int_{0}^{\infty}\frac{d\theta_{1}}{2\pi}\int_{0}^{\theta_1}\frac{d\theta_{2}}{2\pi}...\int_{0}^{\theta_{n-1}}\frac{d\theta_{n}}{2\pi}F_{n}^{\beta\alpha}(\{\theta+i\pi \})e^{imt\sum_{j=1}^{N}\cosh\theta_{j}}\vert\theta_{1},...,\theta_{n}\rangle_{\beta}\\ \nonumber
= & \sum_{n=0}\frac{e^{iE_{\beta}t}}{n!}\int_{-\infty}^{\infty}\frac{d\theta_{1}}{4\pi}\dots\frac{d\theta_{n}}{4\pi}F_{n}^{\beta\alpha}(\theta_{1}+i\pi ,\dots,\theta_{n}+i\pi)e^{imt\sum_{j=1}^{N}\cosh\theta_{j}}\vert\theta_{1},\dots,\theta_{n}\rangle_{\beta}\\
= & e^{iE_{\beta}t}\Bigl(F_{0}^{\beta\alpha}\vert0\rangle_{\beta}+\int_{-\infty}^{\infty}\frac{d\theta}{4\pi}F_{1}^{\beta\alpha}(\theta+i\pi )e^{imt\cosh\theta}\vert\theta\rangle_{\beta}+\dots\Bigr)\,.
\end{align}
We abbreviated  $F_{n}^{\beta\alpha}(\theta_{1}+i\pi ,\dots,\theta_{n}+i\pi)$ by $F_{n}^{\beta\alpha}(\{\theta+i\pi \})$  and we spelled out the contribution of the one-particle states. In the second line we used only the basis of the in-states. In the third line we used the properties of the form factors together with the similar properties of the states to extend the integration domains.

\subsection{Bulk expectation value after the boundary quench}\label{sec:BulkVEVquench}

In the following we monitor what happens in the bulk after the boundary
quench by inserting a field $\Phi(x,t)$ and calculating its vacuum
amplitude 
\begin{equation}
\mathcal{G}_{\beta\alpha}^{\Phi}(x,t)=\ _{\beta}\langle0\vert\Phi(x,t)\psi_{\beta\alpha}(0)\vert0\rangle_{\alpha}\label{eq:amplitude}
\end{equation}
The bulk operator is inserted at $x<0$ and $t>0$ and the boundary condition is changed at $t=0$ from $\alpha$ to $\beta$\footnote{Note, that we simplified the notation compared to~\eqref{eq:Gdef1}, since in this section the left boundary is absent. Although, later in the finite volume BTCSA treatment we will restore the original notation, however the left boundary is always chosen to be $\mathbb I$.}.

\begin{figure}
\begin{centering}
\includegraphics[width=5cm]{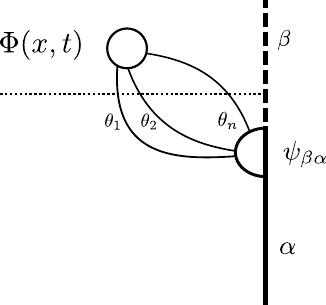}
\par\end{centering}
\caption{The correlator $G_{\beta\alpha}^{\Phi}(x,t)$ 
is calculated by inserting a complete system of multiparticle states with $\beta$ boundary condition at the dashed line. The result is a sum for products of the form factors of the boundary changing operators and the matrix element of the bulk operator between $\beta$ boundary states. \label{fig:BulkBdryform}}
\end{figure}

Let us insert a system of energy eigenstates of the Hilbert space
with $\beta$ boundary condition after the quench, see Figure \ref{fig:BulkBdryform}:
\begin{equation}
\mathcal{G}_{\beta\alpha}^{\Phi}(x,t)=\sum_{\vert\{\theta\}\rangle_{\beta}\in{\cal H}_{\beta}}{}_{\beta}\langle0\vert\Phi(x,t)\vert\{\theta\}\rangle_{\beta}\,_{\beta}\langle\{\theta\}\vert\psi_{\beta\alpha}(0)\vert0\rangle_{\alpha}\,,
\end{equation}
where we lightened the notation by introducing $\vert\{\theta\}\rangle_{\beta}$
for multiparticle states with rapidities $\{\theta_{1},\dots,\theta_{n}\}$. 
The summation for the complete system of states literally means integration for multiparticle in-states, which can be extended for the full unrestricted domain by the reflection and permutation properties of the states and the form factors:
\begin{equation}
\mathcal{G}_{\beta\alpha}^{\Phi}(x,t)=\sum_{n=0}^{\infty}\frac{1}{n!}\int_{-\infty}^{\infty}\frac{d\theta_{1}}{4\pi}\dots\frac{d\theta_{n}}{4\pi}{}_{\beta}\langle0\vert\Phi(x,t)\vert\{\theta \}\rangle_{\beta}\,_{\beta}\langle\{\theta\}\vert\psi_{\beta\alpha}(0)\vert0\rangle_{\alpha}
\end{equation}

In the first factor we use the $\beta$ Hamiltonian to transport
the operator from $0$ to $t$, while in the second we just replace
it with the known boundary form factors to get
\begin{equation}
\mathcal{G}_{\beta\alpha}^{\Phi}(x,t)=\sum_{\vert\{\theta\}\rangle_{\beta}\in{\cal H}_{\beta}}{}_{\beta}\langle0\vert\Phi(x,0)\vert\{\theta\}\rangle_{\beta}F_{n}^{\psi_{\beta\alpha}}(\{\theta+i\pi\})e^{iE(\{\theta\})t}\,.
\end{equation}
where the energy of the multiparticle state is $E(\{\theta\})=E_\beta+\sum_j m\cosh\theta_j $. 
\begin{figure}
\begin{centering}
\includegraphics[width=4cm]{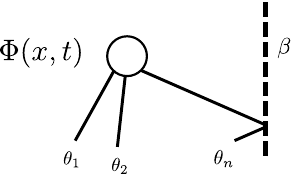}\hspace{2cm}\includegraphics[width=5cm]{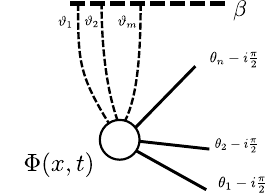}\caption{The matrix elements of the bulk operator $_{\beta}\langle0\vert\Phi(x,0)\vert\theta_{1},\theta_{2},\dots,\theta_{n}\rangle_{\beta}$
on the left and its bulk form factor expansion using the boundary
state on the right. We emphasized that the boundary Hilbert space
contains particles and their reflected companions. \label{fig:rotB}}
\par\end{centering}
\end{figure}

In calculating the matrix element of the bulk operator between boundary
multiparticle states, ${}_{\beta}\langle0\vert\Phi(x,0)\vert\{\theta\}\rangle_\beta$,  we rotate the space-time by $i\pi/2$, see Figure~\ref{fig:rotB}.
 This connects the boundary in space to the boundary in time and replaces the boundary Hilbert space with the bulk Hilbert space and the boundary state. The correspondence between the matrix elements reads as 
\begin{equation}
_{\beta}\langle0\vert\Phi(x,0)\vert\{\theta\}\rangle_{\beta}=\sum_{\{\sigma\}=\pm} R_\beta(\{\sigma \theta\})\langle\{\sigma \theta-\frac{i\pi}{2}\}\vert)\Phi(0,ix)\vert B\rangle_{\beta}\,,
\end{equation}
where we kept in mind that a boundary state $\vert \theta_1,\dots ,\theta_n \rangle_{\beta}$ contains a state and its all possible reflected companion states $\vert \sigma_1\theta_1,\dots ,\sigma_n \theta_n \rangle_{\beta}$ , which come with their generalized reflection factors $R_\beta(\sigma_1 \theta_2,\dots ,\sigma_n \theta_n)$.\footnote{In the case of two particles we have eight terms, some of them are displayed in eq. \eqref{eq:2ptstate}.}  We indicated these factors formally by $R_\beta(\{\sigma \theta\})$. By changing the orders of the particles, $S$-matrix factors also appear, but they are taken care by the permutation property of the  bulk form factors. The bulk form factor on the rhs., however, does not know about the reflection property of the boundary form factor, $ \vert\{\theta\}\rangle_{\beta} $,  so we have to insert the reflected terms by hand.  In the further analyses we keep only one-particle states, when the formula reads as 
\begin{equation}
_{\beta}\langle0\vert\Phi(x,0)\vert\theta\rangle_{\beta}=\Bigl (\bigl \langle\theta-\frac{i\pi}{2}\big\vert+R_\beta(\theta)\bigl \langle-\theta-\frac{i\pi}{2}\big \vert \Bigr)\Phi(0,ix)\vert B\rangle_{\beta}\,.
\end{equation}
This formula connects the matrix elements of a bulk operator between the vacuum and one-particle states in the boundary setting to bulk form factors including the boundary state. This is a very non-trivial relation, which will be checked numerically by TCSA. 

We now calculate these matrix elements by inserting a complete bulk system $\vert\vartheta_{1},\dots,\vartheta_{m}\rangle$. In the generic term we have
\begin{equation}
\langle\{\sigma \theta-\frac{i\pi}{2}\}\vert\Phi(0,ix)\vert B\rangle_{\beta}=\sum_{\vert\{\vartheta\}\rangle\in{\cal H}}\langle\{\sigma \theta-\frac{i\pi}{2}\}\vert\Phi(0,ix)\vert\{\vartheta\}\rangle\langle\{\vartheta\}\vert B\rangle_{\beta}\,.
\end{equation}
Due to the specific form of the boundary state (\ref{eq:Bstate}) only zero momentum states  will contribute to the sum. We then use crossing symmetry and the bulk time evolution operator to obtain 
\begin{equation}
\langle\{\sigma \theta-\frac{i\pi}{2}\}\vert\Phi(0,ix)\vert B\rangle_{\beta}=\sum_{\vert\{\vartheta\}\rangle\in{\cal H}}\langle0\vert\Phi(0,0)\vert\{\vartheta,\sigma \theta+\frac{i\pi}{2}\}\rangle e^{E(\{\vartheta,\sigma \theta+\frac{i\pi}{2}\})x}\langle\{\vartheta\}\vert B\rangle_{\beta}\,.
\end{equation}
where $E(\{\vartheta\})=\sum_j m \cosh \vartheta_j$ is the energy of the multiparticle bulk state. 
This final form includes the bulk form factors $F^{\Phi}(\{\vartheta,\sigma \theta+\frac{i\pi}{2}\})$ and the overlap of the boundary state with the bulk states. 
 
 The bulk form factors take the form \cite{Zamolodchikov:1990bk}
\begin{equation}
F_n^\Phi(\theta_1,\dots ,\theta _n)=\langle \Phi \rangle H_n Q_n(x_1,\dots ,x_n) \prod_{i<j}\frac{f(\theta _i-\theta_j)}{x_i+x_j}
\end{equation}
with $x_i =e^{\theta_i}$. The expectation value is  $   \langle \Phi \rangle =-1.239394325 m^{-2/5} $ and $Q_0=Q_1=1$, while $Q_2=x_1+x_2$.   All higher terms are  known, but we do not need them explicitly here. 

Putting these ingredients together, we arrive at the following all-particle form-factor representation of the post-quench one-point function:
\begin{equation}
    \mathcal{G}_{\beta\alpha}^{\Phi}(x,t)=\sum_{\substack{|\{\vartheta\}\rangle\in{\cal H}, \\  \{\sigma\}=\pm, \\ |\{\theta\}\rangle_{\beta}\in{\cal H}_{\beta}}}R_\beta (\{\sigma \theta\})F^{\Phi}(\{\vartheta,\sigma \theta-\frac{i\pi}{2}\})
\langle\{\vartheta\}\vert B\rangle_{\beta}F^{\psi_{\beta\alpha}}(\{\theta+i\pi \})e^{E(\{\vartheta,\sigma \theta+\frac{i\pi}{2}\})x+iE(\{\theta\})t}
\end{equation}

Here the symbolic sums include the integrations over rapidities, together with the appropriate symmetry factors. 
The set $\{\theta\}$ labels the particles emitted by the boundary-changing operator, while the set $\{\vartheta\}$ labels the particles contained in the boundary state associated with the final boundary condition $\beta$. The signs $\sigma_j=\pm$ encode the two possible reflected images of each particle. The shift by $-i\pi/2$ is the usual crossed kinematics relating the boundary channel to the bulk form factor.

This equation is the central form-factor result of this section. It expresses the real-time response after the boundary quench entirely in terms of on-shell data: bulk form factors of $\Phi$, boundary-changing form factors of $\psi_{\beta\alpha}$, the reflection factor $R_{\beta}$, and the boundary-state amplitudes of the post-quench boundary condition. In this sense the boundary quench problem is reduced to the standard bootstrap data of the massive integrable theory. The price to pay is that, because two complete systems of states are inserted, the resulting expansion has the complexity of a bulk three-point function. This is consistent with the intuitive folded picture: the boundary reflects the excitation, so that the bulk operator effectively couples not only to the particles emitted by the quench, but also to their reflected images.

The expansion is asymptotic in the regime of large time $t$ and large distance from the boundary, $-x\gg m^{-1}$. The latter condition suppresses higher terms in the boundary-state expansion, while the former suppresses multiparticle contributions by stationary phase. Thus, in the regime relevant for the late-time dynamics, the dominant contribution is obtained by retaining only the vacuum term and the one-particle term emitted by the boundary-changing operator. This gives
\begin{equation}
\begin{aligned} \mathcal{G}_{\beta\alpha}^{\Phi}(x,t) &=\,_{\beta}\!\langle 0 \vert \Phi(x,0) \vert 0 \rangle_{\beta}\; {}_{\beta}\!\langle 0 \vert \psi_{\beta\alpha}(0) \vert 0 \rangle_{\alpha} \\ &\quad + \int_{0}^{\infty} \frac{d\theta}{2\pi}\; F^{\alpha\beta}_1(\theta + i\pi)\, e^{i m t \cosh\theta} \\ &\qquad \times \Bigg[ e^{m x \cosh\!\left(\theta - i\frac{\pi}{2}\right)} \left( F_1^\Phi + \frac{g_\beta}{2}\, F_2^\Phi\!\left(\tfrac{i\pi}{2} + \theta, 0\right) e^{m x} + \dots \right) \\ &\qquad\quad +\, e^{m x \cosh\!\left(\theta + i\frac{\pi}{2}\right)} R_\beta(\theta)\, \left( F_1^\Phi + \frac{g_\beta}{2}\, F_2^\Phi\!\left(\tfrac{i\pi}{2} - \theta, 0\right) e^{m x} + \dots \right) \Bigg] + \dots \end{aligned}
\label{eq:leading-ff-expansion}
\end{equation}
The terms proportional to $F_1^\Phi F_1^{\alpha \beta}$ describe the direct exchange of a single particle between the boundary-changing operator and the bulk field. The terms $F_2^\Phi F_1^{\alpha \beta}$ describe the process in which the bulk field also couples to a virtual particle emitted from the boundary state. Since this correction is proportional to $e^{mx}$, it is exponentially suppressed as the operator is moved away from the boundary\footnote{We remind the reader that we defined the theory on the negative half-line, i.e. $x<0$.}. In the numerical regime considered below, already at distances of order $x=-3$ the boundary-coupling contribution is negligible.

The physical content of \eqref{eq:leading-ff-expansion} is transparent. After the quench, the bulk operator relaxes towards the vacuum expectation value corresponding to the new boundary condition, multiplied by the amplitude for changing between the two boundary ground states. The approach to this asymptotic value is not exponential, but oscillatory and algebraic. The oscillation frequency is fixed by the one-particle mass, while the power-law damping follows from the stationary point of the relativistic dispersion relation at zero rapidity.

Indeed, the large-$t$ (fixed $x$) behaviour of the integral in \eqref{eq:leading-ff-expansion} is governed by the small-rapidity region. Expanding around $\theta=0$, one would generically obtain a leading contribution proportional to
\[
\frac{e^{i(mt+\pi/4)}}{\sqrt{t}}\,
F^{\alpha\beta}_{1}(i\pi)\,
\bigl(1+R_{\beta}(0)\bigr)
\left[
F_{1}^{\Phi}
+
\frac{g_{\beta}}{2}e^{mx}
F_{2}^{\Phi}\!\left(\frac{i\pi}{2},0\right)
+\cdots
\right].
\]
However, for the integrable boundary conditions considered here the reflection factor satisfies
\(
R_{\beta}(0)=-1 .
\)
Consequently, the leading stationary-phase contribution cancels between the direct and reflected terms. The first non-vanishing term is therefore one order lower in the stationary-phase expansion, and the late-time correction behaves as
\begin{equation}
\mathcal{G}_{\beta\alpha}^{\Phi}(x,t)
=
{}_{\beta}\!\langle 0|\Phi(x,0)|0\rangle_{\beta}\,
{}_{\beta}\!\langle 0|\psi_{\beta\alpha}(0)|0\rangle_{\alpha}
+
\frac{A_{\beta\alpha}^{\Phi}(x)}{t}\,e^{imt}
+\cdots .
\label{eq:late-time-asymptotics}
\end{equation}
where $A_{\beta\alpha}^\Phi(x)$ is obtained from the subleading term of the $\theta $ expansion of the phase and the integrand. Thus the boundary reflection changes the leading decay from the generic $t^{-1/2}$ behaviour to $t^{-1}$. This cancellation is a direct dynamical consequence of the boundary: the particle emitted by the boundary-changing operator interferes destructively with its reflected image at zero rapidity.  

So far we assumed that $x$ is kept fixed and $t\to \infty$, so that $x/t\to 0$. If, however, $x$ and $t$ are comparable then in the asymptotic formula we need to take the $t\to \sqrt{t^2-x^2}$ changes to get a better approximation.

In the following section, we test the prediction \eqref{eq:late-time-asymptotics}, and in particular the characteristic $e^{imt}/t$ relaxation, using a truncated conformal space approach adapted to the present boundary-quench protocol.

\section{TCSA confirmations}\label{sec:TCSA}

In this section, we test the form-factor predictions derived above within the TCSA framework. We first explain how the method is adapted to the present boundary-quench setup. As a check of the implementation, we recall the computation of the finite-volume energy spectrum and eigenstates for the $\mathbb{I}\mathbb{I}$ and $\mathbb{I}\phi$ boundary conditions.

Let us emphasize that the Lee--Yang model is perturbed only in the bulk, while the conformal boundary conditions are kept fixed. After mapping the strip to the upper half-plane, the Hamiltonian therefore takes the same formal form in both cases,
\begin{equation}
\label{eq:massiveHam}
H=
\frac{\pi}{L}
\left[
L_{0}-\frac{c}{24}
+
\left(\frac{mL}{\pi\kappa}\right)^{2-2h}
\int_{0}^{\pi}
\Phi(e^{i\theta},e^{-i\theta})\,d\theta
\right] .
\end{equation}
The difference between the two boundary conditions enters through the corresponding Hilbert spaces and through the conformal matrix elements of the bulk field. We describe these ingredients below.

\subsection{TCSA on the strip with $\mathbb{II}$ boundary conditions}

We start by analysing the spectrum of the scaling Lee--Yang model on the strip with identity boundary condition on both sides. Since in the boundary setting there is only one copy of the Virasoro algebra, the bulk field formally can be written as a product of chiral primary fields: $\Phi(z,\bar{z})\sim\varphi(z)\varphi(\bar{z}).$ Additionally, the Hilbert space
is built over the identity module with the Virasoro generators as 
\begin{equation}
{\cal H}_{\mathbb{II}}=V_{0}=\{L_{-n_{1}}\dots L_{-n_{m}}|0\rangle;\quad n_{m}>1\quad,\quad n_{i}>n_{i+1}+1\}\,.
\end{equation}
Here, the specific selection rules guarantee that there are no singular vectors \cite{Feigin:1991wv,Bajnok:2014fca}. We truncate this Hilbert space at a given energy level  $\Lambda
$. In calculating the matrix elements of $\Phi(z,\bar{z})$ we need its expectation value on the UHP
\begin{equation}
\langle0|\Phi(z,\bar{z})|0\rangle=B_{\mathbb{I},\mathbb{I}}(i(\bar{z}-z))^{\frac{2}{5}}\,,
\end{equation}
together with its commutation relations with the Virasoro modes. Keeping in mind that $\Phi(z,\bar{z})$ can be represented as a product of its chiral components $\varphi(z)\varphi(\bar{z})$ we merely need to use the commutation relation 
\begin{equation}
[L_{n},\varphi(z)]=z^{n}([L_{0},\varphi(z)]+nh\varphi(z))\,.
\end{equation}
We presented these relations in a specific form including $L_{0}$, which can be evaluated to the left or to the right, except those stuck between the two chiral halfs. These stuck modes at the end can be turned to differential operators $L_{0}^{n}\to(2h+\bar{z}\partial_{\bar{z}})^{n}$, which act on the expectation value, i.e. on $(z-\bar{z})^{-2h}.$ We then have to integrate the obtained expression with $z=e^{i\theta},\bar{z}=e^{-i\theta}$ from $0$ to $\pi$, which we evaluated numerically. 

In finite, but large volume, we expect a linear volume dependence of the energy spectrum, due to the bulk energy density $\epsilon_\text{bulk}$. Moreover, each boundary contributes to the energy with $\epsilon_{\mathbb{I}}$,  thus the ground state energy is expected to have the following volume dependence \cite{Dorey:1997yg}:
\begin{equation}\label{eq:BY11}
E_{0}(L)=\epsilon_{\mathrm{bulk}}L+2\epsilon_{\mathbb{I}}+O(e^{-mL})\quad;\quad\epsilon_{\mathrm{bulk}}=-\frac{1}{4\sqrt{3}}\quad;\quad\epsilon_{\mathbb{I}}=\frac{1}{4}(\sqrt{3}-1)\,.
\end{equation}

States above the ground state are one- or multi-particle states, subject to the so-called Bethe--Yang quantization condition, which reads for one-particle states as:
\begin{equation}
2mL\sinh\theta_{n}-2i\log R_{\mathbb{I}}(\theta_{n})=2\pi n\quad;\quad E_{n}-E_{0}(L)=m\cosh\theta_{n}\,,
\end{equation}
where $n\in \mathbf{N}$ and the energy relative to the ground state $E_{n}-E_{0}(L)$ are given in terms of the quantized rapidities. We neglegected vacuum polarization effects of order $O(e^{-mL})$. 

\begin{figure}

\includegraphics[width=0.45\textwidth]{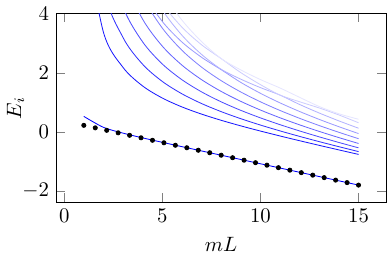}\hspace{1cm}\includegraphics[width=0.45\textwidth]{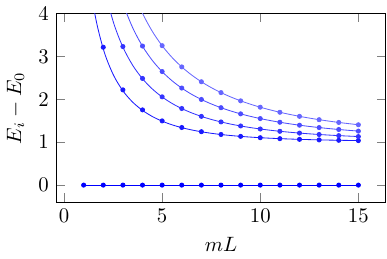}

\caption{We present the low lying energy spectrum as function of the dimensionless volume of the scaling Lee-Yang model on the strip with $\mathbb{II}$ boundary conditions on the left (blue lines, shading indicates the different energy levels), together with $\epsilon_{\mathrm{bulk}}L+2\epsilon_{\mathbb{I}}$ (black dots). The difference between the one-particle energies and the vacuum is compared to the Bethe-Yang lines in the right panel (TCSA data is denoted by dots, while the corresponding solutions to the one-particle Bethe--Yang equations are indicated straight lines).}
\label{fig:spec11}
\end{figure}

We obtained the dimensionless volume dependence of the spectrum with TCSA at level cut-off $\Lambda=16$. The results are presented in Figure~\ref{fig:spec11}. In the left panel, we indicated~\eqref{eq:BY11} and the one-particle Bethe--Yang spectrum on the right.

\subsection{TCSA on the strip with $\mathbb{I\phi}$ boundary conditions}

We can repeat the same analysis with boundary conditions $\mathbb{I}$
and $\phi$ on the two sides of the strip. The Hilbert space consists
of modes 
\begin{equation}
{\cal H}_{\mathbb{I}\phi}=V_{h}=\{L_{-n_{1}}\dots L_{-n_{m}}\vert h\rangle;\quad n_{m}>0\quad,\quad n_{i}>n_{i+1}+1\}\,.
\end{equation}

The calculation is very similar to the one we had before. We first
determine the vacuum expectation value (VEV) of the field from conformal covariance
\begin{equation}
\langle \psi_{\phi\mathbb{I}}\vert\Phi(z,\bar{z})\vert\psi_{\phi\mathbb{I}}\rangle=d_1(z-\bar{z})^{\frac{1}{5}}\bar{z}^{\frac{1}{5}}F_{2,1}\Bigl(\frac{1}{5},\frac{2}{5};\frac{4}{5}\Big \vert\frac{\bar{z}}{z}\Bigr)+d_2(z-\bar{z})^{\frac{1}{5}}\frac{\bar{z}^{\frac{2}{5}}}{z^{\frac{1}{5}}}F_{2,1}\Bigl(\frac{2}{5},\frac{3}{5};\frac{6}{5}\Big \vert\frac{\bar{z}}{z}\Bigr)\,.
\end{equation}
This is the same solution we analyzed before, but adapted to the situation, when the boundary changing operator is at the origin (and at infinity). We have the following analytic expressions
\begin{equation}
d_1=-\frac{\sqrt{2-\frac{4}{\sqrt{5}}+2i\sqrt{1-\frac{2}{\sqrt{5}}}}\pi\Gamma\left(\frac{2}{5}\right)\Gamma\left(\frac{6}{5}\right)}{\Gamma\left(\frac{4}{5}\right)^{2}\left(\Gamma\left(\frac{1}{5}\right)\Gamma\left(\frac{4}{5}\right)-\Gamma\left(\frac{2}{5}\right)\Gamma\left(\frac{3}{5}\right)\right)}\quad;\quad d_2=\frac{(-1)^{2/5}\sqrt{2-\frac{4}{\sqrt{5}}+2i\sqrt{1-\frac{2}{\sqrt{5}}}}\pi}{\Gamma\left(\frac{1}{5}\right)\Gamma\left(\frac{4}{5}\right)-\Gamma\left(\frac{2}{5}\right)\Gamma\left(\frac{3}{5}\right)}\,,
\end{equation}
which numerically give $d_1=-1.54628-1.12344i$ and $d_2=-0.393076+1.20976i$. We then commute the Virasoro modes through $\Phi(z,\bar{z})\sim\varphi(z)\varphi(\bar{z})$ and replace the stuck $L_{0}$ modes with the differential operators, which eventually act on the expectation value. We then integrate numerically the terms. The numerical truncated Hamiltonian turns out to be real, which is a non-trivial consistency check. 

\begin{figure}

\includegraphics[width=0.45\textwidth]{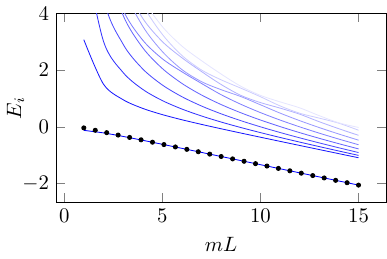}\hspace{1cm}\includegraphics[width=0.45\textwidth]{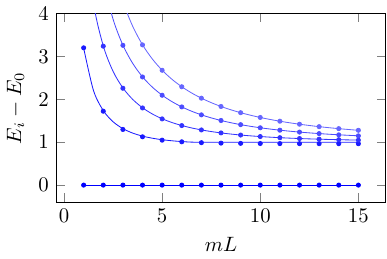}

\caption{The low lying energy spectrum with $\mathbb{I\phi}$ boundary conditions
on the left, together with $\epsilon_{\mathrm{bulk}}L+\epsilon_{\mathbb{I}}+\epsilon_{\phi}$.
The difference between the one-particle energies and the vacuum is
compared to the Bethe-Yang lines on the right. \label{fig:spec1phi}}
\end{figure}

The spectrum obtained at level cut-off $16$ can be seen in Figure \ref{fig:spec1phi} on the left, together with the one particle Bethe--Yang spectrum on the right.

On the left we indicated again the bulk energy and boundary energy
contributions to the groundstate
\begin{equation}
E_{0}(L)=\epsilon_{\mathrm{bulk}}L+\epsilon_{\mathbb{I}}+\epsilon_{\phi}+O(e^{-mL})\quad;\quad\epsilon_{\mathbb{\phi}}=\epsilon_{\mathbb{I}}+\sin\frac{\pi b}{6}\,,
\end{equation}
where, in the absence of a boundary perturbation, we have to take $b=-\frac{1}{2}$. On the right we subtracted these contributions and displayed the solution of the one-particle Bethe-Yang energies, fixed by
\begin{equation}\label{eq:BY1phi}
2mL\sinh\theta_{n}-i\log R_{\mathbb{I}}(\theta_{n})-i\log R_{\phi}(\theta_{n})=2\pi n\quad;\quad E_{n}-E_0(L)=m\cosh\theta_{n}
\end{equation}
where vacuum polarization effects are neglected as in the previous case. 
Clearly, we have a complete understanding of the large volume  spectrum.

\subsection{Form factors of the boundary field}

We now test the form factors of the boundary operators. We place a
chiral boundary field $\varphi(0)$ on the $\phi$ boundary and calculate
its matrix elements between the vacuum and one-particle states. Although
this analysis was already performed in \cite{Kormos:2007qx}, we repeat
it in our framework to check the consistency of our results. The vacuum
expectation value in infinite volume is 
\begin{equation}
_{\phi}\langle0\vert\varphi\vert0\rangle_{\phi}=-\frac{5}{6h_{\mathrm{crit}}}\frac{\cos\frac{\pi b}{6}}{\cos(\frac{\pi}{10}(2b+1))}m^{-\frac{1}{5}}\quad;\quad h_{\mathrm{crit}}=0.68529\,.
\end{equation}
To calculate the matrix elements of the boundary field $\varphi$, we use its commutation relation with the Virasoro modes. We then normalize it with its three-point function  \eqref{eq:bobo}.
\begin{figure}
\includegraphics[width=0.45\textwidth]{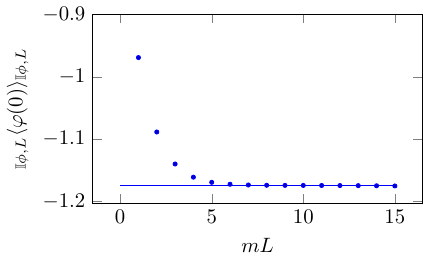}\hspace{1cm}\includegraphics[width=0.45\textwidth]{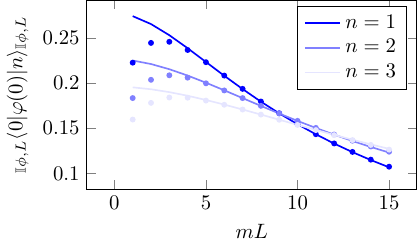}

\caption{The vacuum expectation value of the boundary field $\varphi$ on the
strip with boundary conditions $\mathbb{I}\phi$, on the left, while its low-lying one-particle form factors on the right as the function of the volume. One-particle states are determined using eq.~\eqref{eq:BY1phi} and $|n\rangle_{\mathbb{I}\phi,L}$ denotes the one-particle states corresponding to different Bethe--Yang quantum numbers. Dots represent the TCSA data, while straight lines are the corresponding theoretical finite volume approximation~\eqref{eq:FFfinitevol}.}
 \label{fig:varphi}
\end{figure}
With this normalization, we recover the boundary VEV, as we demonstrate in the left panel of Figure~\ref{fig:varphi}.

We now turn to the one-particle form factors. The infinite volume one-particle boundary form factors take the form \cite{Bajnok:2006ze} 
\begin{equation}
F_{1}(\theta)={}_{\phi}\langle0\vert\varphi\vert\theta\rangle_{\phi}={}_{\phi}\langle0\vert\varphi\vert0\rangle_{\phi}\ H_{1} r(\theta)
\end{equation}
and 
\begin{equation}\label{eq:rint}
r(\theta)=\frac{i\sinh\theta\quad\exp\left\{ \int_{0}^{\infty}\frac{dt}{t}\left[\frac{1}{\sinh\frac{t}{2}}-2\cosh\frac{t}{2}\cos[\frac{t}{\pi}(\frac{i\pi}{2}-\theta)]\frac{\sinh\frac{5t}{6}+\sinh\frac{t}{2}-\sinh\frac{t}{3}}{\sinh^{2}t}\right]\right\} }{(\sinh\theta-i\sin\frac{\pi(b+1)}{6})(\sinh\theta-i\sin\frac{\pi(b-1)}{6})}
\end{equation}
To compare the numerically "measured" matrix elements, we also have to take into account that finite volume states and infinite volume states are normalized differently \cite{Kormos:2007qx}
\begin{equation}\label{eq:FFfinitevol}
_{\mathbb{I}\phi,L}\langle0|\varphi|n\rangle_{\mathbb{I}\phi,L}=\frac{F_{1}(\theta_n)}{\sqrt{\rho_{1}(\theta_n,L)}}\quad;\quad\rho_{1}(\theta_n,L)=2mL\cosh\theta_n-i\partial_{\theta}(\log R_{\mathbb{I}}(\theta)+\log R_{\phi}(\theta))|_{\theta=\theta_n}\,.
\end{equation}
where we denoted the finite volume ground state and one-particle states by $|0\rangle_{\mathbb{I}\phi,L}$ and $|n\rangle_{\mathbb{I}\phi,L}$. 
We present the results in the right panel of Figure~\ref{fig:varphi} showing excellent agreement. For small volumes, the deviation originates from finite size effects\footnote{In fact, eq.~\eqref{eq:FFfinitevol} only amounts to the $\propto \frac{1}{L}$ corrections, however, there exist exponentially suppressed corrections due to vacuum polarization effects, see~\cite{2008NuPhB.788..167P,2008NuPhB.788..209P}. }, while in large volume, the TCSA method is no longer reliable.

\subsection{Form factors of boundary changing operators}

We now turn to the investigation of the form factors of boundary changing operators \cite{Bajnok:2015iwa}. We calculate the matrix elements of $\psi_{\mathbb{I}\phi}$ and $\psi_{\phi\mathbb{I}}$, which change the boundary conditions, and has matrix elements between the Hilbert spaces ${\cal H}_{\mathbb{II}}$ and ${\cal H}_{\mathbb{I}\phi}$. Let us start with the vacuum-vacuum matrix element. We found numerically that 
\begin{equation}
_{\mathbb{I}\mathbb{I}}\langle0|\psi_{\mathbb{I}\phi}|0\rangle_{\mathbb{I}\phi}=\ {}_{\mathbb{I}\phi}\langle0|\psi_{\phi\mathbb{I}}|0\rangle_{\mathbb{I}\mathbb{I}}\approx0.96158
\end{equation}

We then turned to the finite volume form factors. At large volumes, when vacuum polarization effects are neglected, the form factors are the same as in infinite volume,
\begin{equation}
_{\mathbb{I}\mathbb{I},L}\langle0|\psi_{\phi\mathbb{I}}|n\rangle_{\mathbb{I}\phi,L}=\frac{F_{1}^{\phi\mathbb{I}}(\theta_n)}{\sqrt{\rho_{1}(\theta_n,L)}}+O(e^{-mL})\,
\end{equation}
 only the states has to be renormalized with the one-particle density of states
\begin{equation}
\rho_{1}(\theta_n,L)=2mL\cosh\theta_n-2i\partial_{\theta}\log R_{\mathbb{I}}(\theta)|_{\theta=\theta_n}
\end{equation}and
\begin{equation}
F_{1}^{\phi\mathbb{I}}(\theta)={}_{\mathbb{I}}\langle0\vert\psi_{\phi\mathbb{I}}\vert0\rangle_{\phi}H_{1}r^{\mathbb{II}}(\theta)\frac{r_{\phi}(i\pi-\theta)}{2\cosh\theta}\,
\end{equation}
is the form factor of the boundary changing operator in the semi-infinite strip.

Similarly,
\begin{equation}
_{\mathbb{I}\phi,L}\langle0|\psi_{\mathbb{I}\phi}|n\rangle_{\mathbb{I}\mathbb{I},L}=\frac{F_{1}^{\mathbb{I}\phi}(\theta_n)}{\sqrt{\rho_{1}(\theta_n,L)}}+O(e^{-mL})\,,
\end{equation}
with the semi-infinite volume one-particle form factor
\begin{equation}
F_{1}^{\phi\mathbb{I}}(\theta)={}_{\phi}\langle0\vert\psi_{\mathbb{I}\phi}\vert0\rangle_{\mathbb{I}}H_{1}r^{\mathbb{II}}(\theta)\frac{r_{\phi}(\theta)}{2\cosh\theta}\,.
\end{equation}
In the numerical evaluation, we used the functional relation 
\begin{equation}
r_{\phi}(\theta)r_{\phi}(\theta+i\pi)=\frac{1}{4(\cosh\frac{i\pi(3-b)}{6}-\cosh(\theta-\frac{i\pi}{3}))(\cosh\frac{i\pi(3-b)}{6}-\cosh(\theta+\frac{i\pi}{3}))}\,,
\end{equation}
in order to analytically continue the integral representation in~\eqref{eq:rint}.

\begin{figure}
\includegraphics[width=0.45\textwidth]{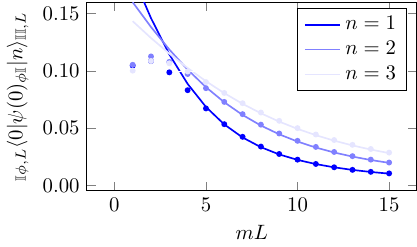}\hspace{1cm}\includegraphics[width=0.45\textwidth]{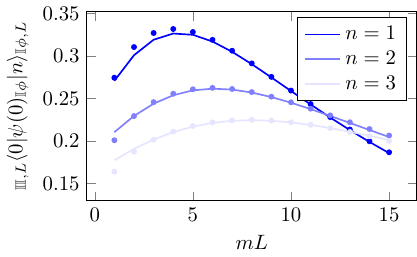}

\caption{Comparison of the finite volume one-particle form factors with the
TCSA result. The operator $\psi_{\phi\mathbb{I}}$ is on the left, while $\psi_{\mathbb{I}\phi}$ is on the right.}
\label{fig:psibff}

\end{figure}

The results with the nice agreement at larger volumes are displayed in Figure~\ref{fig:psibff}, $\psi_{\mathbb{I}\phi}$ on the left, while $\psi_{\phi\mathbb{I}}$ on the right.

\subsection{Form factors of the bulk field}\label{sec:bulkFF}

\begin{figure}

\includegraphics[width=0.45\textwidth]{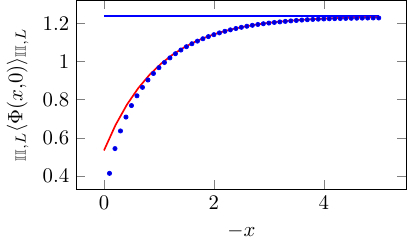}\hspace{1cm}\includegraphics[width=0.45\textwidth]{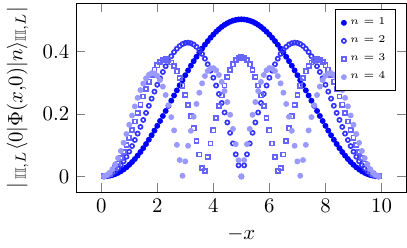}

\caption{Left panel: VEV of the bulk field $\Phi(x,0)$ on the strip
with identity boundary conditions. Dots represent the TCSA data, while blue line is the bulk expectation value and the red one contains the correction coming from a standing one-particle state. Right panel: we present the TCSA results for the absolute value of the first four one-particle form factors $_{\mathbb{I}\mathbb{I},L}\langle0|\Phi(x,0)|n\rangle_{\mathbb{I}\mathbb{I},L}$. The data were computed in a volume $L=10$.
\label{fig:bulkvev11}}
\end{figure}

To compute the amplitude \eqref{eq:amplitude} in the form factor approach, we need the matrix element of the bulk field between boundary multi-particle states. As a first step we evaluate the expectation value of the bulk field with boundary condition $\mathbb{I}$ on both sides
\begin{equation}
_{\mathbb{I}\mathbb{I}}\langle0|\Phi(x,t)|0\rangle_{\mathbb{I}\mathbb{I}}=\langle0\vert\Phi(z,\bar{z})\vert0\rangle\Bigl(\frac{\pi}{L}z\frac{\pi}{L}\bar{z}\Bigr)^{-\frac{1}{5}}\quad;\quad z=e^{\frac{i\pi}{L}(x-t)}\quad;\quad\bar{z}=e^{-\frac{i\pi}{L}(x+t)}
\end{equation}
As the matrix elements of $\Phi(z,\bar{z})$ between the states of the identity module have been already obtained to construct the Hamiltonian, we can easily evaluate the expectation value in the massive case. From a boundary state calculation \cite{Dorey:2000eh} it is expected to behave as 
\begin{equation}
_{\mathbb{I}}\langle0\vert\Phi(x,t)\vert0\rangle_{\mathbb{I}}=\langle0\vert\Phi(t,ix)\vert B\rangle=\langle\Phi\rangle(1+\frac{g}{2}F^\Phi_{1}e^{mx}+\dots)\,,
\end{equation}
which is indicated in the left panel of Figure~\ref{fig:bulkvev11}. The solid lines represent the infinite volume VEV and the inclusion of the one-particle form factor term, while the dots represent the TCSA data.

Next, we extend these results for excited states. The measured overlaps for one-particle states $_{\mathbb{I}\mathbb{I},L}\langle0|\Phi(x,0)|n\rangle_{\mathbb{I}\mathbb{I},L}$ are shown in the figure on the right. 

\begin{figure}
\begin{centering}
\includegraphics[width=0.45\textwidth]{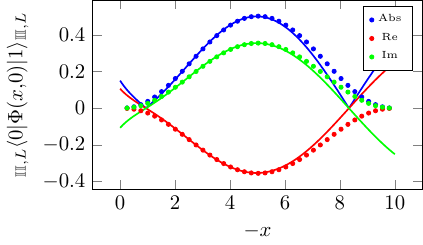}\hspace{1cm}\includegraphics[width=0.45\textwidth]{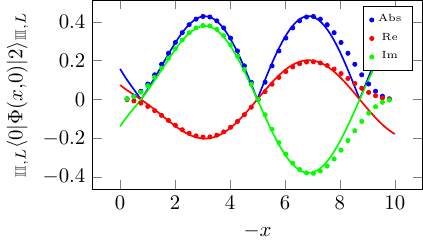}
\par\end{centering}
\caption{We present the TCSA values of $\vert _{\mathbb{I}}\langle0\vert\Phi(x,0)\vert\theta\rangle_{\mathbb{I}}\vert $ with dots for the first (left) and second (right) one particle states. Solid lines are (\ref{eq:bulkvev1pt}). The data presented were computed in a volume $L=10$. 
}\label{fig:Phi1pt} 
\end{figure}

Let us understand these matrix elements from the semi-infinite setting. We rotate the space-time to use the boundary state again 
\begin{align}
\nonumber _{\mathbb{I}}\langle0|\Phi(x,0)|\theta\rangle_{\mathbb{I}}=&\left(\langle\theta-i\frac{\pi}{2}|+R(\theta)\langle-\theta-i\frac{\pi}{2}|\right)\Phi(0,ix)\vert B\rangle_{\mathbb{I}}\\=&F_{1}\left(e^{mx\cosh(\theta-i\frac{\pi}{2})}+R(\theta)e^{mx\cosh(-\theta-i\frac{\pi}{2})}\right)+\dots\label{eq:bulkvev1pt}
\end{align}
where we note that the boundary one-particle state $\vert\theta\rangle_{\mathbb{I}}$
can be formally identified with $\vert\theta\rangle+R(\theta)\vert-\theta\rangle$,
i.e. it behaves as $\vert\theta\rangle_{\mathbb{I}}=R(\theta)\vert-\theta\rangle_{\mathbb{I}}$ and this is the way we have to represent it in the bulk after the space-time interchange. Note, that in the second line we kept only the ground state contribution to the boundary state.
We can go one order higher in the particle number by evaluating 
\begin{equation}\label{eq:1pttobulk}
\langle\theta-i\frac{\pi}{2}|\Phi(0,ix)\left(|0\rangle+\frac{g}{2}|\theta=0\rangle+\dots\right)=e^{mx\cosh(\theta-i\frac{\pi}{2})}\left(F_{1}+\frac{g}{2}F_{2}(\theta+\frac{i\pi}{2},0)e^{-mx}+\dots\right)
\end{equation}
In Figure \ref{fig:Phi1pt} we show these corrections for the first two finite volume states. The results clearly indicates that the crossing of $_{\mathbb{I}}\langle0\vert\Phi(x,0)\vert\theta\rangle_{\mathbb{I}}$ includes both the state and its reflected counterpart. 

\subsection{Vacuum amplitude of the bulk field after the quench}

In the following we investigate the vacuum amplitude of the bulk field 
\begin{equation}
\mathcal{G}_{\mathbb{I}\phi}^{\Phi}(x,t)= \,_{\mathbb{I}}\langle0\vert\Phi(x,t)\psi_{\mathbb{I}\phi}(0)\vert0\rangle_{\phi}
\end{equation}
in semi-infinite volume\footnote{The left boundary is thought to be $\mathbb I$, and pushed to $-\infty$.}. By recalling the methodology in Subsection of~\ref{sec:BulkVEVquench}, we write
\begin{equation}
\mathcal{G}_{\mathbb{I}\phi}^{\Phi}(x,t)=\,_{\mathbb{I}}\langle0\vert e^{itH_{\mathbb{I}}}\Phi(x,0)e^{-itH_{\mathbb{I}}}\psi_{\mathbb{I}\phi}\vert0\rangle_{\phi}\,,
\end{equation}
with $H_\mathbb{I}$ being the Hamiltonian on the semi-infinite plane with boundary condition $\mathbb I$.
Keeping only the vacuum and one-particle terms we get
\begin{equation}\label{eq:Tdepfinal}
\mathcal{G}_{\mathbb{I}\phi}^{\Phi}(x,t)=\,_{\mathbb{I}}\langle\Phi(x,0)\rangle_{\mathbb{I}}\ _{\mathbb{I}}\langle\psi_{\mathbb{I}\phi}\rangle_{\phi}+\int_{0}^{\infty}\frac{d\theta}{2\pi}{}_{\mathbb{I}}\langle0\vert\Phi(x,0)\vert\theta\rangle_{\mathbb{I}}\ _{\mathbb{I}}\langle\theta\vert\psi_{\mathbb{I}\phi}\vert0\rangle_{\Phi}e^{-imt\cosh\theta}\,.
\end{equation}
Based on the discussion in Subsection~\ref{sec:bulkFF}, sufficiently far from the boundaries the leading one-particle matrix element takes the form
\begin{equation}
{}_{\mathbb{I}}\langle 0|\Phi(x,0)|\theta\rangle_{\mathbb{I}}
=
F_{1}^{\Phi}
\left[
e^{m x\cosh\left(\theta-\frac{i\pi}{2}\right)}
+
R_{\mathbb I}(\theta)\,
e^{m x\cosh\left(-\theta-\frac{i\pi}{2}\right)}
\right]
+\cdots .
\end{equation}
Keeping only this leading contribution, the oscillating part of the post-quench one-point function is
\begin{equation}
\begin{aligned}
\mathcal{G}_{\mathbb I\phi}^{\Phi}(x,t)\big|_{\rm osc}
&=
{}_{\mathbb I}\langle0|\psi_{\mathbb I\phi}|0\rangle_{\phi}\,
H_{1}\,F_{1}^{\Phi}
\int_{0}^{\infty}\frac{d\theta}{2\pi}\,
\left[
e^{-i m x\sinh\theta}
+
R_{\mathbb I}(\theta)\,
e^{i m x\sinh\theta}
\right]
\\
&\hspace{25mm}\times
e^{-i m t\cosh\theta}\,
\left(r^{\mathbb I\mathbb I}(\theta)\right)^{*}
\frac{\left(r_{\phi}(i\pi-\theta)\right)^{*}}{2\cosh\theta}
+\cdots .
\end{aligned}
\label{eq:Tdepfinal}
\end{equation}
The integrand is symmetric under the appropriate continuation to negative rapidities, so that the expression may equivalently be viewed as a folded version of the usual full-line one-particle contribution. We have also evaluated the contribution of two-particle intermediate states. In the parameter range where the comparison with TCSA is performed, these corrections are numerically negligible. We therefore focus on the one-particle expression~\eqref{eq:Tdepfinal} and ask how its space-time dependence is reproduced by the truncated Hamiltonian calculation.

In the TCSA calculation we work at fixed strip width $L$, with left boundary condition $\mathbb I$. The time-dependent correlation function is represented as
\begin{equation}
\mathcal{G}_{\mathbb I\phi}^{\Phi}(x,t)
=
{}_{\mathbb I\mathbb I,L}\langle0|
e^{i t H_{\mathbb I\mathbb I}}
\Phi(x,0)
e^{-i t H_{\mathbb I\mathbb I}}
\psi_{\mathbb I\phi}
|0\rangle_{\mathbb I\phi,L}.
\end{equation}
Operationally, we first act with the truncated matrix representation of the boundary-changing operator on the finite-volume ground state in the $\mathbb I\phi$ Hilbert space. This produces a state in the post-quench Hilbert space $\mathcal H_{\mathbb I\mathbb I}$. We then evolve this state with the truncated Hamiltonian $H_{\mathbb I\mathbb I}$ and evaluate the matrix element of the bulk field in the same Hilbert space. In practice, the calculation is performed in a finite conformal basis. Since this basis is not orthonormal, all matrix products are accompanied by the corresponding Gram matrices; further details are given in Appendix~\ref{app:TCSA}.

\begin{figure}
    \centering
    \includegraphics[width=0.8\linewidth]{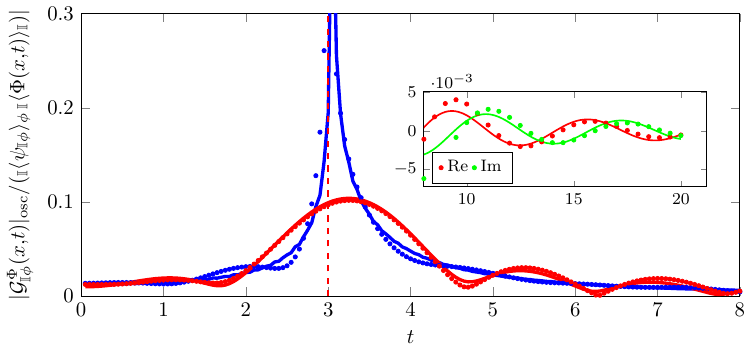}
    \caption{Vacuum-vacuum matrix element vs. TCSA for $L=10$ and $x=-3$. The straight blue curve is the total one-particle contribution of our formula, while the red one is integrated up to rapidities below a cut-off compatible with the truncation in TCSA. The red dots show the result of the TCSA calculation at truncation level 16, while the blue ones contain the first four counter-terms, see Appendix~\ref{app:TCSAcutoff}. We also plot the real and imaginary parts of the expression~\eqref{eq:Tdepfinal}, together with the function $-0.0229711e^{i t}(t^2-3^2)^{-1/2}$ which is the leading asymptotic behaviour, see the discussion in the end of~\ref{sec:BulkVEVquench}.}
    \label{fig:tcsa1}
\end{figure}

\begin{figure}
    \centering
    \includegraphics[width=0.8\linewidth]{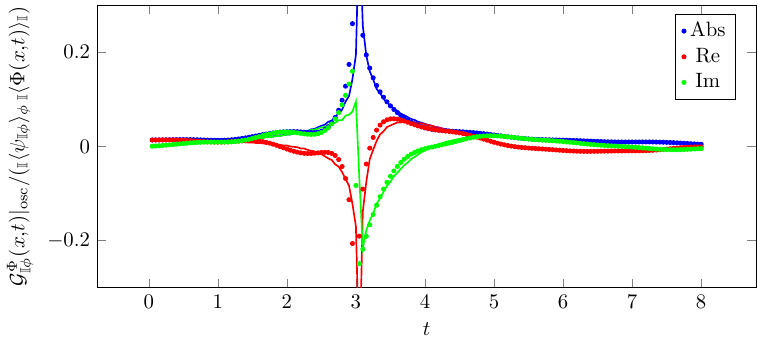}
    \caption{Modulus, real and imaginary part of the oscillating contribution compared to renormalized TCSA data for $L=10$ and $x=-3$.} 
    \label{fig:tcsaReIm}
\end{figure}

A direct comparison between the raw TCSA data and the continuum expression,presented in Figure~\ref{fig:tcsa1}, reveals an important cutoff effect. The unprocessed TCSA signal displays strong finite-cutoff oscillations and only a modest light-cone peak, whereas the continuum one-particle expression predicts a much more pronounced enhancement near the light cone. The origin of this discrepancy is that TCSA retains only a finite set of intermediate states. Although Eq.~\eqref{eq:Tdepfinal} contains only one-particle contributions, it integrates over the full one-particle energy spectrum. The sharp light-cone structure is therefore sensitive to high-energy rapidities, which are precisely the modes most affected by the truncation. We also present the real- and imaginary parts separately in Figure~\ref{fig:tcsaReIm}.

To make this comparison more transparent, we also truncate the rapidity integral in Eq.~\eqref{eq:Tdepfinal}, restricting it to low-energy one-particle states in analogy with the TCSA cutoff. The resulting expression exhibits oscillations very similar to those observed in the raw TCSA data. This confirms that the absence of a sharp light-cone peak at finite cutoff is not a failure of the form-factor prediction, but rather a manifestation of the ultraviolet truncation. After applying the regularization procedure described in Appendix~\ref{app:TCSAcutoff}, the TCSA data is well approximated by the full one-particle contribution.

\begin{figure}
    \centering
    \includegraphics[width=0.8\textwidth]{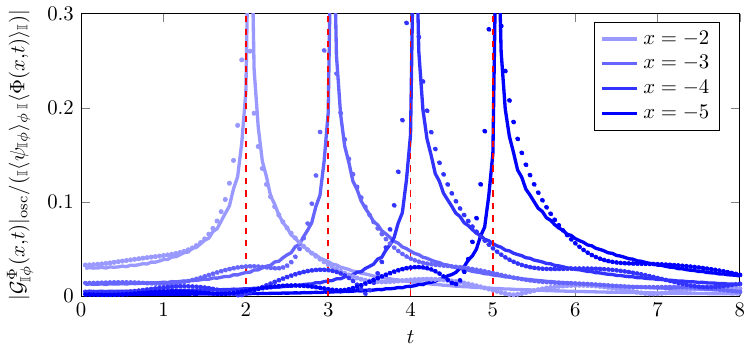}
    \caption{Time evolution of the bulk one-point function at various space locations. The one-particle form factor approximations~\eqref{eq:Tdepfinal} are indicated by straight lines, while renormalized TCSA data is presented with dots. We see qualitative agreement of the two approximations.}
    \label{fig:Massivequench}
\end{figure}

Further comparisons are shown in Fig.~\ref{fig:Massivequench} for several values of the bulk insertion point. The agreement demonstrates that the two approaches capture the same physical effect: after the boundary condition is changed, information propagates from the boundary into the bulk, producing a clear light-cone spreading pattern.

Let us summarize our findings. We developed a double-expansion for the time evolution of one-point functions after an integrable boundary-changing quench. The boundary condition in the initial state was incorporated through the boundary-state formalism after Wick rotation, while an additional bulk resolution of the identity was inserted to express the time evolution in terms of bulk and boundary-changing form factors. In practice, both expansions were truncated to their first non-trivial contributions: we retained at most one-particle states in the boundary and bulk channels. We also evaluated the leading two-particle corrections and found them to be numerically negligible in the parameter regime used for the comparison. This provides an a posteriori justification for the one-particle approximation in the regime considered here.

The same quantity was computed independently using a renormalized TCSA approach. In this calculation, the cutoff dependence was modeled using exact CFT data, motivated by the observation in Appendix~\ref{app:TCSAcutoff} that the leading cutoff dependence is largely insensitive to whether the time evolution is generated by the massive Hamiltonian or by its ultraviolet conformal counterpart. The agreement between the form-factor prediction and the renormalized TCSA data is remarkably precise.

We also clarified the origin of the apparent discrepancy between the raw finite-cutoff TCSA data and the continuum form-factor expression. The difference is caused by the absence of high-energy degrees of freedom in the truncated Hilbert space. Indeed, by imposing an analogous finite-rapidity cutoff on the one-particle form-factor integral, we reproduced the characteristic finite-cutoff TCSA signal. This demonstrates that the sharp light-cone structure of the continuum result is built from high-energy one-particle modes and is therefore particularly sensitive to truncation effects.

These results provide a non-trivial verification of the form-factor description of integrable boundary-changing quenches. At the same time, they give a more detailed understanding of cutoff effects in Hamiltonian truncation methods. This is important beyond the integrable setting: while form-factor techniques rely on exact solvability, Hamiltonian truncation can be applied to generic quantum field theories. The present comparison therefore provides both a benchmark for the integrable calculation and a guide for future applications of renormalized truncation methods to more general non-equilibrium boundary dynamics.

\section{Conclusion and outlook}\label{sec:conc}

In this work, we proposed a boundary-quench protocol in which the boundary condition of the initial state is changed instantaneously by the insertion of a boundary-condition-changing operator. We studied the resulting vacuum-to-vacuum matrix elements of bulk fields between the pre- and post-quench ground states.

In the conformal case, we showed that, for specific quenches, the problem reduces to the computation of chiral four-point functions. These can be decomposed into chiral conformal blocks, with coefficients fixed by OPE constraints. In the massive integrable case, we developed a double form-factor expansion, combining boundary-state techniques with a bulk resolution of the identity. We used the Lee--Yang model as a concrete testing ground, where the required conformal data and bulk and boundary form factors are available. In the massive theory, we compared the form-factor predictions with a renormalized truncated conformal space approach calculation and found convincing agreement. In addition, we proposed a method for computing matrix elements of bulk fields between boundary multiparticle states and checked it against TCSA for a set of one-particle states.

In analytically tractable quantum-field-theory quenches, late-time one-point functions typically approach their asymptotic value through oscillations with a model- and protocol-dependent envelope. The oscillation frequencies are set either by one-particle masses~\cite{2014JPhA...47N2001D,2017JPhA...50h4004D,sineGquench2014JSMTE..10..035B,2018ScPP....5...27H,CastroAlvaredoetal2020E8,KiralyLencses2025arXiv250619596K} or by twice these masses~\cite{2012JSMTE..04..017S,IsingEnt}, depending on the particle content of the initial state. One-particle components, generated by a nonzero one-particle form factor of the quench operator, lead to single-particle frequencies, whereas quasiparticle-pair components and nonzero two-particle form factors produce double frequencies. The envelope likewise varies across settings, ranging from persistent oscillations~\cite{2014JPhA...47N2001D,2017JPhA...50h4004D,CastroAlvaredoetal2020E8}, through power-law decay~\cite{2011PhRvL.106v7203C,IsingEnt}, to exponential damping~\cite{2012JSMTE..04..017S,sineGquench2014JSMTE..10..035B,IsingEnt}. Our results fit naturally into this picture. The quench locally generates one-particle excitations which propagate inside the light cone and leave behind damped oscillations after the light cone has passed. For inhomogeneous quenches, one would generically expect a stationary-phase envelope of order $t^{-1/2}$~\cite{2020NuPhB.95415002D}. In our integrable boundary-changing protocol, however, the leading contribution cancels due to reflection, so the dominant late-time behaviour is $t^{-1}$. We expect this to be the generic asymptotic form for quenches to integrable boundary conditions with $R(0)=-1$, provided both the boundary-changing operator and the measured operator have nonzero one-particle form factors.

The Lee--Yang results demonstrate that our framework provides a viable computational approach to boundary-changing quenches. It would be very interesting to extend the analysis to other, unitary integrable models, such as the Ising and three-state Potts models. This would require determining the necessary bulk form factors, boundary form factors, and boundary-changing form factors. In the Ising case, existing results for free fermions~\cite{Lesage:1998hh} could provide a useful starting point. The conformal analysis should also admit a natural generalization. More realistic models typically contain several primary fields, and therefore offer a richer set of possible boundary-changing quenches. Although the corresponding conformal computations would involve more complicated combinations of chiral blocks, they remain tractable in principle.

Another promising direction is to study spin-chain realizations of these boundary-quench protocols. In such systems, tensor-network-based algorithms could be used to simulate the real-time dynamics, in analogy with recent applications to global quenches~\cite{CastroAlvaredoetal2020E8,KiralyLencses2025arXiv250619596K}. This would provide an independent numerical approach and could help bridge the gap between continuum field-theory predictions and lattice realizations.

Finally, boundary-changing quenches may also be relevant for experimental and quantum-simulation platforms. In particular, spin chains implemented on digital~\cite{VovroshKnolle2021NatSR..1111577V,Roy2024NatCo..15.5901L,Knolle22025arXiv251202516H,Igloi2025arXiv251203341H} or analogue quantum computers~\cite{Falsevac2025arXiv251204637C} could offer a direct route to observing the light-cone propagation and relaxation phenomena described in this work.


\acknowledgments

 The authors acknowledge funding from the Ministry of Culture and Innovation and the National Research, Development and Innovation Office (NKFIH) through the OTKA Grant NKKP-152467. ML was supported by the Bolyai J\'anos Research Scholarship of the Hungarian Academy of Sciences.

\appendix
\section{Truncation details}\label{app:TCSA}

\subsection{Computation of correlators in TCSA}

To approximate the time dependent correlator~\eqref{eq:amplitude} in a massive perturbation of a conformal field theory, we use the TCSA approach. Let us recall that the truncation works in finite volume, while our form factor approximation in Section~\ref{sec:massive} was devised in a semi-infinite volume setting. However, close to one of the boundaries (say the right one), we expect the effect of the other boundary to be negligible. To this end, in the TCSA we used a relatively large volume, and we set identity boundary condition on the left boundary.

Let us introduce the following notations for the matrix elements of the boundary changing operator, the bulk operator and the Gramm-matrix of the Virasoro representation at hand:
\begin{align}
    C_{\alpha \beta, ij} &= \,_{\mathbb{I}\alpha}\langle i |\psi_{\alpha \beta} | j\rangle_{\mathbb I \beta}\,, \\
    B_{ij}(z,\bar z) &= \,_{\mathbb I\mathbb I}\langle i | \Phi(z,\bar z) | j \rangle_{\mathbb I \mathbb I}\,, \\
    M_{\alpha \beta,ij} &= \,_{\alpha \beta} \langle i | j \rangle_{\alpha \beta}\;.
\end{align}

Fortunately, all of these matrices have to be computed in order to diagonalize the Hamiltonian, therefore no additional computations are necessary.

Note that, the working Virasoro basis is not orthonormal, therefore in the actual numerical implementation of the Hamiltonian, we have to correct for this. The matrix elements of the Hamiltonian~\eqref{eq:massiveHam} are taken in the Virasoro basis, which is not orthogonal. In the diagonalization procedure this has to be compensated, since diagonalization routines use the action of the operator, not the matrix-elements between the non-orthogonal states.

Finally, the initial state before the boundary change, can be written as the truncated linear combination of BCFT states
\begin{equation}
    \ket{0}_{\mathbb I \phi} =\sum_{i\in \mathcal{H}_{\mathbb I \phi}| N_i<\Lambda} a_i \ket{i}_{\mathbb I \phi}\,,
\end{equation}
where $N_i$ is the descendant level of the state $|i\rangle_{\mathbb I \phi}$. Similarly, the ground state of the post-quench theory is written as
\begin{equation}
    \ket{0}_{\mathbb I \mathbb I} =\sum_{i\in \mathcal{H}_{\mathbb I \mathbb I}| N_i<\Lambda} b_i \ket{i}_{\mathbb I \mathbb I}\,.
\end{equation}
Rememeber, that the Virasoro basis is not orthonormal, therefore these factors are normalised such that
\begin{align}
    \nonumber a^*_i [M_{\mathbb I \phi}]_{ij} a_j & =  1\,, \\
    \nonumber b^*_i [M_{\mathbb I \mathbb I}]_{ij} b_j &=  1\,, \\
\end{align}
where the repeated indices are summed over.
The truncated TCSA approximation of the correlator is then computed as:
\begin{equation}
G_{\mathbb{I}\phi}^{\Phi}(x,t) \approx \left( \frac{\pi}{L}\right)^{-\frac{3}{5}} 
b_j^* \left[ M_{\mathbb I \mathbb I}\right]_{jk} \left[e^{i H_{\mathbb I \mathbb I}t}\right]_{k l} \left[ M^{-1}_{\mathbb I \mathbb I}\right]_{lm} B_{mn}(e^{- \frac{\pi}{L}i x},e^{\frac{\pi}{L}i x})\left[e^{-i H_{\mathbb I \mathbb I}t}\right]_{np} \left[ M^{-1}_{\mathbb I \mathbb I}\right]_{pq} \left[C_{\mathbb I \phi}\right]_{qr}a_r
\end{equation}
where we compensated for the non-orthogonality of the working basis by the appropriate insertions of $M$ and its inverse.

\subsection{Cut-off dependence}\label{app:TCSAcutoff}

In this subsection we review the cut-off dependence of correlators, first in pure CFT then in the massive perturbation.

\paragraph{CFT Correlator}

Let us recall the result for the quench amplitude in the CFT, after quenching the boundary from $\phi$ to the identity:

\begin{equation}\label{eq:appendix4pt}
G_{\mathbb I, \mathbb I \phi}^{\Phi}(z,\bar{z})\sim\langle0\vert\psi_{\mathbb I\phi}(1)\Phi(z,\bar{z})\psi_{\phi\mathbb I}(0)\vert0\rangle\sim((1-z)(1-\bar{z})(z-\bar{z})z\bar{z})^{-\frac{2h}{3}}f\Bigl(\frac{(1-z)\bar{z}}{(1-\bar{z})z}\Bigr)
\end{equation}
with coefficients
\begin{equation}
c_{2}=\frac{\sqrt{\frac{6}{\sqrt{5}}-2}\pi}{\Gamma(\frac{1}{5})\Gamma(\frac{4}{5})-\Gamma(\frac{2}{5})\Gamma(\frac{3}{5})}\quad;\quad c_{1}=-\frac{\Gamma(\frac{2}{5})\Gamma(\frac{6}{5})}{\Gamma(\frac{4}{5})^{2}}c_{2}
\end{equation}

In the TCSA we compute a truncated version of this function:

\begin{equation}
    \sum_{n,m}\,_{\mathbb I \mathbb I}\langle 0|\Phi\left(e^{-i\frac{\pi}{L}x},e^{i\frac{\pi}{L}x}\right) |n \rangle_{\mathbb I \mathbb{I}}\,_{\mathbb I \mathbb{I}}\langle n | z^{L_0}|m \rangle_{\mathbb I \mathbb{I}}\,_{\mathbb I \mathbb{I}} \langle m | \psi_{\mathbb I \phi}(1)|0\rangle_{\mathbb I \phi}
\end{equation}

\begin{figure}
    \centering
    \includegraphics[width=0.8\linewidth]{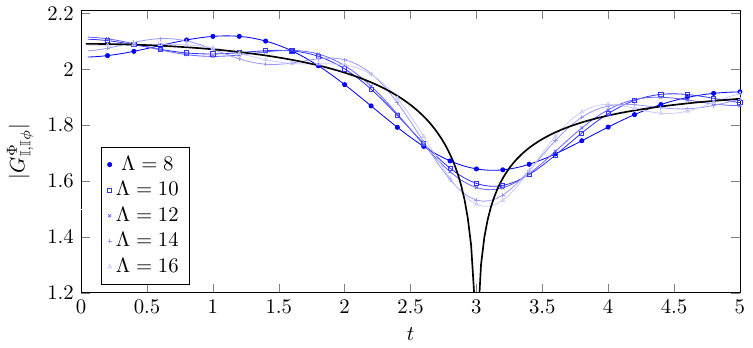}
    \caption{$G_{\mathbb{I},\phi}^{\Phi}$ for different cut-offs for $L=10$ and $x=-3$. TCSA for various cut-offs reproduce the expansion~\eqref{eq:Gexp}, truncated at the same cut-off. The exact results is presented with a solid black line.}
    \label{fig:CFTcutoff}
\end{figure}

One can introduce the following variables:
\begin{equation}
    Z = z;\qquad X=\frac{z}{\bar{z}}
\end{equation}
and expand around $Z=\infty$ with fixed $X$ and truncate the expansion to get the approximation:
\begin{equation}\label{eq:Gexp}
     G_{\mathbb I,\mathbb I \phi}^{\Phi, (\Lambda)}(Z,X) = \sum_{n=0}^\Lambda g_{\mathbb I \phi,n}^{\Phi}(X) Z^n.
\end{equation}
By construction TCSA at truncation level $N$ calculates exactly the observable in this expansion upto the corresponding cut-off. This is demonstrated for several cut-offs on Figure ~\ref{fig:CFTcutoff}. The expansion converges only very slowly to the exact results.
This allows us to construct the exact result, by simply subtracting the expansion~\eqref{eq:Gexp} from the TCSA result and adding back the exact result, to get the renormalized TCSA results:

\begin{equation}
     G_{\mathbb I \phi,\text{TCSA,ren}}^{\Phi, (\Lambda)}(z,\bar{z}) = G_{\mathbb I \phi,\text{TCSA}}^{\Phi, (\Lambda)}(z,\bar{z}) -  G_{\mathbb I \phi}^{\Phi, (\Lambda)}(Z,X) + G_{\mathbb I \phi}^{\Phi}(z,\bar{z})\,.
\end{equation}
This, of course results in the exact CFT correlation function. In fact, we were able to carry this "approxiamation" because the exact correlator is known analytically. Moreover, correlators between descendant states i.e.
\begin{equation}
    \langle0|\prod_i L_{n_i}\psi_{\mathbb I\phi}(1)\Phi(z,\bar{z})\prod_k L_{-n_k}\psi_{\phi\mathbb I}(0)|0\rangle\,,
\end{equation}
where various products of Virasoro generators are inserted can be approximated using TCSA in the same way: the descendant correalors can be derived from the original one by applying certain differential operators and the same expansion in $z$ can be carried out. We computed the truncated contributions for correlators between the first descendant in the Hilbert space $\mathcal{H}_{\mathbb{I}\phi}$ and the second in $\mathcal{H}_{\mathbb{I}\mathbb{I}}$, and checked numerically that TCSA computes the corresponding truncation of the expansions.

The situation is different in the massive case, however, as we present in the next subsection, these results can be used to reduce the truncation effects.

\paragraph{Massive case}

In the massive case there are two sources of error. One is in the approximation of the ground state caused by truncation. This however, seems to be negligible: indeed, form factors computed from this approximation show a mild truncation dependence. Another, and more important source of truncation effect is the truncation in the internal summations, as it already happens in the conformal case.

The upshot of the previous discussion is that by truncating the internal channel in the evaluation of a conformal correlator one introduces a cut-off dependence, which can be exactly modelled for CFT objects.

In the massive case the pre-quench approximate state in the TCSA is given as a linear combination of states with boundary conditions $\gamma$ and  $\alpha$:
\begin{equation}
    \ket{0}_{\gamma\alpha} =\sum_{n\in \mathcal{H}_{\gamma,\alpha}|E_n<\Lambda} a_n \ket{n}_{\gamma\alpha}\,,
\end{equation}
while the post-quench ground state is approximated in the $\gamma\beta$ Hilbert-space:
\begin{equation}
    _{\gamma\beta}\bra{0} = \sum_{m\in \mathcal{H}_{\gamma\beta}|E_m<\Lambda} b^*_m \,_{\gamma\beta}\bra{m}\,.
\end{equation}

\begin{figure}
    \centering
    \includegraphics[width=0.8\linewidth]{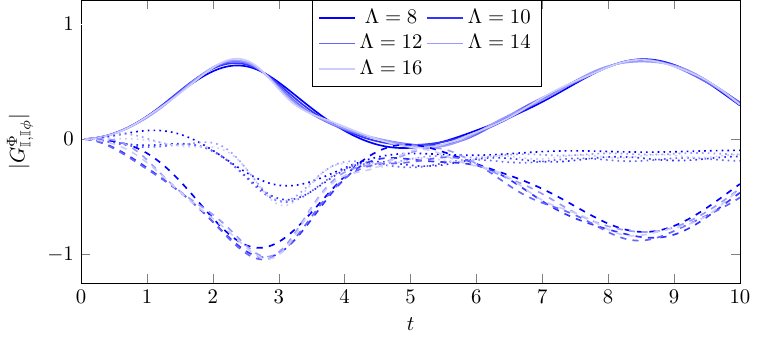}
    \caption{Cut off dependence of the BCFT four-point function. Dotted/dashed lines are generated using conformal/massive time evolution ($t=0$ values are subtracted). Both have a characteristic cut-off dependent oscillations, the leading is absent when the difference of the two quantities is taken (solid lines). This signifies that the cause of such oscillation is the truncation of the internal summations. The large oscillation of the massive time evolved case is due to the presence of the energy scale $m$ in the perturbed theory: the period is roughly $2\pi$, in agreement with our choice of the coupling such that $m=1$.}
    \label{fig:4ptdiff}
\end{figure}

The TCSA approximation of the vacuum-vacuum matrix element then amounts to the evaluation of the objects:
\begin{equation}
    _{\gamma,\beta}\bra{m} \tilde\Phi(z,\bar{z}) \psi_{\beta,\alpha}(1) \ket{n}_{\gamma,\alpha}\,,
\end{equation}
where the tilde means that the field has to be time evolved using the massive Hamiltonian, hence the above objects are not BCFT correlators. However, we can still approximate their cut-off dependence by using the BCFT version:
\begin{equation}\label{eq:bcftfunction}
    _{\gamma,\beta}\bra{m} \Phi(z,\bar{z}) \psi_{\beta,\alpha}(1) \ket{n}_{\gamma,\alpha}\,,
\end{equation}
supposing that the leading cut-off dependence is caused by the internal truncation. Indeed, this is the case, as we demonstrate in Figure~\ref{fig:4ptdiff} for the ground state.

We can then carry out an approximate renormalization of the massive TCSA time evolution results by using the renormalized version of the CFT objects~\eqref{eq:bcftfunction}. In fact, the most important contributions come from the contribution of the highest weight state and its first descendant to the initial state and the highest weight and its second descendant to the post-quench ground state. Carrying out the expansions for the correlators for higher cut-offs is a cumbersome task, due to variety of branch cuts, hence we used the TCSA truncated version of the four point functions in the renormalization process. Finally, our partially "renormalized" approximation of the time dependence is the following:
\begin{equation}
    \mathcal{G}^{\Phi,(\Lambda)}_{\mathbb{I}\phi,\text{ren}}(x,t) = \mathcal{G}^{\Phi,(\Lambda)}_{\mathbb{I}\phi,\text{TCSA}}(x,t) + \sum_{i,j=1,2} b_i^{(\Lambda)*} a^{(\Lambda)}_j \mathcal{C}_{ij}(x,t)
\end{equation}
where the counter-terms read as:

\begin{eqnarray}
    \mathcal C_{1,1}^{(\Lambda)} &=& \,_{\mathbb{I},\mathbb{I}}\bra{0} \Phi(x,t) \psi_{\mathbb{I},\phi}(0,0) \ket{0}_{\mathbb{I},\phi} - \,_{\mathbb{I},\mathbb{I}}\bra{0} \Phi(x,t) \psi_{\mathbb{I},\phi}(0,0) \ket{0}_{\mathbb{I},\phi}^{(\Lambda)}\\
    \mathcal C_{1,2}^{(\Lambda)} &=& \,_{\mathbb{I},\mathbb{I}}\bra{0} \Phi(x,t) \psi_{\mathbb{I},\phi}(0,0) \ket{L_{-1}}_{\mathbb{I},\phi} - \,_{\mathbb{I},\mathbb{I}}\bra{0} \Phi(x,t) \psi_{\mathbb{I},\phi}(0,0) \ket{L_{-1}}_{\mathbb{I},\phi}^{(\Lambda)}\\
    \mathcal C_{2,1}^{(\Lambda)} &=& \,_{\mathbb{I},\mathbb{I}}\bra{L_{-2}} \Phi(x,t) \psi_{\mathbb{I},\phi}(0,0) \ket{0}_{\mathbb{I},\phi} - \,_{\mathbb{I},\mathbb{I}}\bra{L_{-2}} \Phi(x,t) \psi_{\mathbb{I},\phi}(0,0) \ket{0}_{\mathbb{I},\phi}^{(\Lambda)}\\
    \mathcal C_{2,2}^{(\Lambda)} &=& \,_{\mathbb{I},\mathbb{I}}\bra{L_{-2}} \Phi(x,t) \psi_{\mathbb{I},\phi}(0,0) \ket{L_{-1}}_{\mathbb{I},\phi} - \,_{\mathbb{I},\mathbb{I}}\bra{L_{-2}} \Phi(x,t) \psi_{\mathbb{I},\phi}(0,0) \ket{L_{-1}}_{\mathbb{I},\phi}^{(\Lambda)}\\ 
\end{eqnarray}
and the cut-off dependent conformal four-point functions and vector coefficients ($a^{(\Lambda)},b^{(\Lambda)}$) were extracted from TCSA.

\bibliographystyle{JHEP}
\bibliography{biblio.bib}

\end{document}